\documentclass[aip,apl,twocolumn,reprint]{revtex4-2}
\usepackage{longtable}
\usepackage{morefloats}
\usepackage[dvips]{graphicx}
\usepackage{color}
\usepackage{epsfig,graphicx,amsfonts,amsbsy}
\usepackage{amsmath,amsfonts,amsthm,amssymb}
\usepackage{appendix}
\usepackage{makeidx}
\usepackage{url}
\usepackage{verbatim}
\usepackage[bookmarksnumbered,pdfpagelabels=true,plainpages=false,colorlinks=true,linkcolor=blue,citecolor=blue,urlcolor=blue]{hyperref}
\usepackage[rightcaption]{sidecap}
\usepackage{array}
\usepackage{booktabs}
\usepackage{multirow}
\usepackage{floatrow}
\newlength{\mywidth}
\usepackage{cancel,soul,ulem}

\begin{document}

\title{Magneto-optical detection of spin-orbit torque vector with first-order Kerr effects}
\author{C. Gonzalez-Fuentes}
\affiliation{Instituto de F\'{i}sica, Pontificia Universidad Cat\'{o}lica de Chile, Vicu\~{n}a Mackena 4860, 7820436 Santiago, Chile}
\email{cgonzalef@uc.cl}
\author{M. Abellan}
\affiliation{Centro Cient\'{i}fico y Tecnol\'{o}gico de Valpara\'{i}so - CCTVal,  Universidad T\'{e}cnica Federico Santa Mar\'{i}a, Avenida Espa\~{n}a 1680, 2390123 Valpara\'{i}so, Chile}
\affiliation{Departamento de Física, Universidad Técnica Federico Santa María, Valparaíso 2390123, Chile}
\author{S. Oyarzún}
\affiliation{Departamento de Física, Universidad de Santiago de Chile, Santiago 9170124, Chile}
\author{C. Orellana}
\affiliation{Centro Cient\'{i}fico y Tecnol\'{o}gico de Valpara\'{i}so - CCTVal,  Universidad T\'{e}cnica Federico Santa Mar\'{i}a, Avenida Espa\~{n}a 1680, 2390123 Valpara\'{i}so, Chile}
\affiliation{Departamento de Física, Universidad Técnica Federico Santa María, Valparaíso 2390123, Chile}
\date{\today }

\begin{abstract}
We have developed a novel, compact and cost-effective magneto-optical method for quantifying the spin-orbit torque (SOT) effective field vector ($\mathbf{h_{\text{SO}}}$) in magnetic thin films subjected to spin current injection. The damping-like ($\mathbf{h}^{\text{DL}}_{\text{SO}}$ ) component of the vector is obtained by the polar Kerr response arising from the out-of-plane magnetization tilting, whereas the field-like component ($\mathbf{h}^{\text{FL}}_{\text{SO}}$) is obtained by current-induced hysteresis-loop shifting study, using conventional longitudinal Kerr magnetometry.

We tested our method in bilayers comprising NiFe, CoFeB (ferromagnetic layers) and Pt, Pd, Ta (non-magnetic layers). Our findings revealed a damping-like SOT efficiency $\xi^{\text{DL}}$ of $0.089 \pm 0.006$, $0.019 \pm 0.002$, and $-0.132 \pm 0.009$ for Pt, Pd, and Ta, respectively, very in line with the most accepted values for those materials. The $\mathbf{h}^{\text{FL}}_{\text{SO}}$/$\mathbf{h}^{\text{DL}}_{\text{SO}}$ ratio was $0.35 \pm 0.02$ for NiFe/Pt and $0.14 \pm 0.02$ for Ta/CoFeB bilayers when the ferromagnetic layer thickness is 4 nm.

A key advancement over the state-of-art magneto-optical methods is the use of an oblique light incidence angle, which allows switching between measuring modes for $\mathbf{h}^{\text{DL}}_{\text{SO}}$ and $\mathbf{h}^{\text{FL}}_{\text{SO}}$, without altering the experimental setup. Moreover, our approach relies exclusively on first-order Kerr effects, thereby ensuring its broad applicability to any type of ferromagnetic material.
\end{abstract}

\maketitle
The inter-conversion between charge and spin currents has been a central subject of 
research in condensed matter physics since its prediction \cite{Dyakonov71} and the first experimental
confirmation \cite{Kato04}.
When spin-polarized currents diffuse into a ferromagnetic (FM) material, they induce spin-orbit torques (SOTs) \cite{Manchon19}:  a powerful mechanism for magnetic order manipulation, with applications in magneto-resistive random access memory (MRAM) \cite{Bhatti17}, logic devices and neuro-morphing computing \cite{Torrejon17}.

Although techniques to characterize SOTs have existed for more than a decade \cite{Saitoh06,Pi10},
the emergence of new phenomena such as rotated-symmetry \cite{Amin18,Baek18}, single-layer SOTs \cite{Luo19} or orbital currents\cite{Sala22}, has made more complex the interpretation of the results given by the state-of-art techniques 
for probing SOTs.

To date, these techniques have consisted almost exclusively
on measuring {\it{electrical}} voltages arising from the device under
excitation. In this line, techniques such as second harmonic generation (SHG)
\cite{Kim13,Garello13,Avci14,Avci14b,Woo14,Nguyen16,Nguyen16b,Aykol16,Ou16b} , spin transfer torque ferromagnetic resonance (ST-FMR)
and spin pumping (SP) generated voltages \cite{Saitoh06}, have dominated the scene. 

The techniques mentioned above have a series of problems and limitations in their applicability.
To begin with, in all the {\it{electrical}} detection methods, the signals are potentially contaminated by thermoelectric voltages, which are not always easy to disentangle from the signal of interest \cite{Hayashi14,Kondou16,Vliestra14,Liu21}.

On the other hand, in ST-FMR and SHG, the electrical current injected to excite magnetization dynamics is also necessary for its detection; hence,  the sensitivity critically depends on the fraction
of the electrical current flowing through the FM layer of the structure.  For this reason, these methods are not suitable, in a general way,  for detecting SOTs on FM insulators \cite{Montazeri15}.
Also, in metallic FM/NM bilayers, the sensitivity reduces as the FM layer is too thin or resistive compared to the NM layer. In the opposite scenario, a thin and highly resistive NM layer sacrifices the strength of the SHE.

Another drawback of all the  {\it{electrical}} detection methods is that a complete detection of the SOT vector relies on the anisotropic magneto-resistance (AMR) coefficient of the FM layer, which presents a large variability among 3d-group ferromagnetic alloys. For example, the room temperature AMR coefficient can vary from 1.4\% in NiFe to 0.06\% in CoFeB \cite{McGuire75}. 

In this scenario, the magneto-optical detection of SOTs \cite{Montazeri15,Fan16,Kim19,Celik19} arose as a promising solution to the abovementioned issues, as the magnetization dynamics can be probed without electrical contacts. Pioneering works of Montazeri \cite{Montazeri15,Fan16} showed the detection of damping-like (DL) and field-like (FL) components of SOT on FM layers by controlling the polarization of the incident light.
More recently, Xin. et al. \cite{Xing20} showed the detection of DL and FL effective fields but OPA and IP samples, respectively.

Despite these advances, magneto-optical detection of SOTs is very far from being a widely adopted method for detecting SOTs \cite{Nguyen21}, as {\it{electrical}} methods are.
This may be in part because most of the setups employed to detect SOTs use normal-light incidence angle, a configuration convenient for probing DL-SOT but unpractical for the obtention of in-plane magnetization components and, hence, for exploring FL-SOT
\cite{Celik19,Xing20}.
In addition, a reliable magneto-optical characterization of the SOT vector has been demonstrated only via quadratic MOKE (Q-MOKE), a second-order effect, which is strongly dependent on the polarization of the incident light but also on crystallographic symmetries of the FM material \cite{Postava02,Silver19}. This complexifies the characterization of single crystal materials such as magnetic iron garnets. In addition, combining MOKE with Q-MOKE requires changes in the experimental setup when moving from DL-SOT to FL-SOT characterization.


This work presents a magneto-optical method for detecting DL and FL-SOT effective fields, namely $\mathbf{h}^{\text{DL}}_{\text{SO}}$ and $\mathbf{h}^{\text{FL}}_{\text{SO}}$, via first-order, polar, and longitudinal MOKE, respectively. 
The experimental setup is straightforward, as is commonly employed for longitudinal MOKE magnetometry with nearly-crossed polarizers \cite{Qiu00}, with the additional requirement of the polarizers rotation control. ln summary, our setup is the same as the generalized magneto-optical ellipsometry technique\cite{Berger97,Gonzalez-Fuentes19}.
Moreover, our method does not require fabrication-intensive samples, and both components of the SOT vector can be studied without modifying the setup. Very importantly, our method relies only on first-order Kerr effects, whose magnitude scales directly with the saturation magnetization of the FM layer \cite{Higo18}. Hence, broad applicability to FM materials is guaranteed.

 \begin{figure}[ht]
\includegraphics[width=8cm]{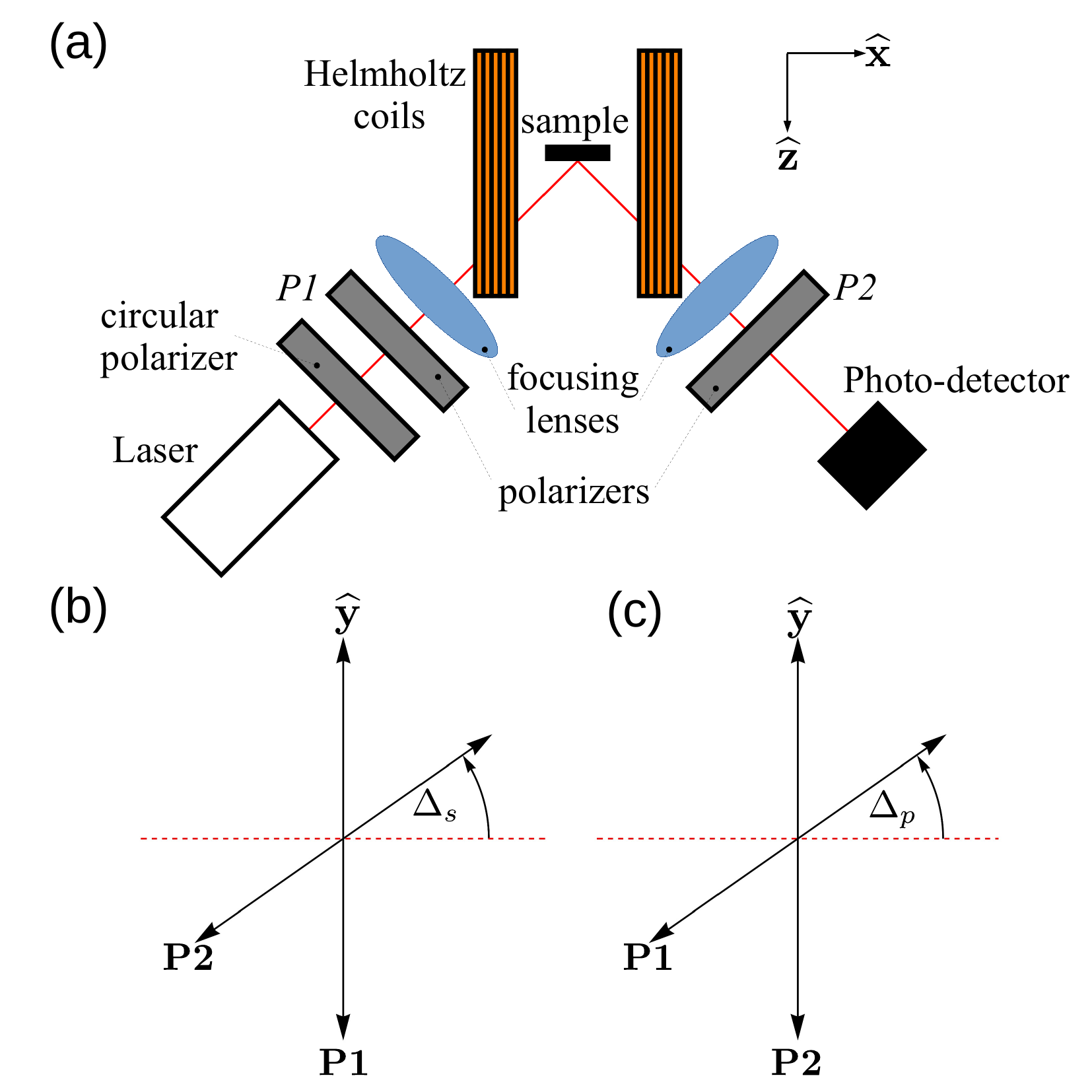}
\caption{(a) Schematic diagram of the magneto-optical ellipsometer.  (b) and (c): system of reference for defining the perturbation angles $\Delta_{s}$ and $\Delta_{p}$ in the {\it{s}} and {\it{p}} measurement mode, respectively. The red dashed line denotes the plane parallel to the optical table.
}
\label{fig1}
\end{figure}

We tested our method in a set of UHV-sputtered FM/NM bilayers: NiFe(8)/Pt(6), NiFe(4)/Pt(6), Ta(8)CoFeB(4), Ta(8)/CoFeB(8) and NiFe(8)/Pd(12), where NiFe =Ni$_{81}$Fe$_{19}$ and CoFeB=Co$_{40}$Fe$_{40}$B$_{20}$. The sequence of materials is described from the bottom to the top layer with the number inside parenthesis indicating the thickness in nm of each layer. Also, a 2 nm Ta capping layer was added to CoFeB-on-top samples as a protective layer against oxidation, which is assumed to be wholly oxidized upon exposition to air.
  
The two samples with $t_{\text{FM}}=\text{4 nm}$ were designed specifically for testing $\mathbf{h}^{\text{FL}}_{\text{SO}}$ detection. In the rest of the samples, we consider $\mathbf{h}^{\text{FL}}_{\text{SO}}$ to be negligible in comparison with the Oersted field\cite{Soya21,Dutta21}. 
All the samples were patterned into 2$\times$1 mm stripes and contacted at the extremes by Au gold pads.

\begin{center}
 \begin{figure*}[ht]
\includegraphics[width=14cm]{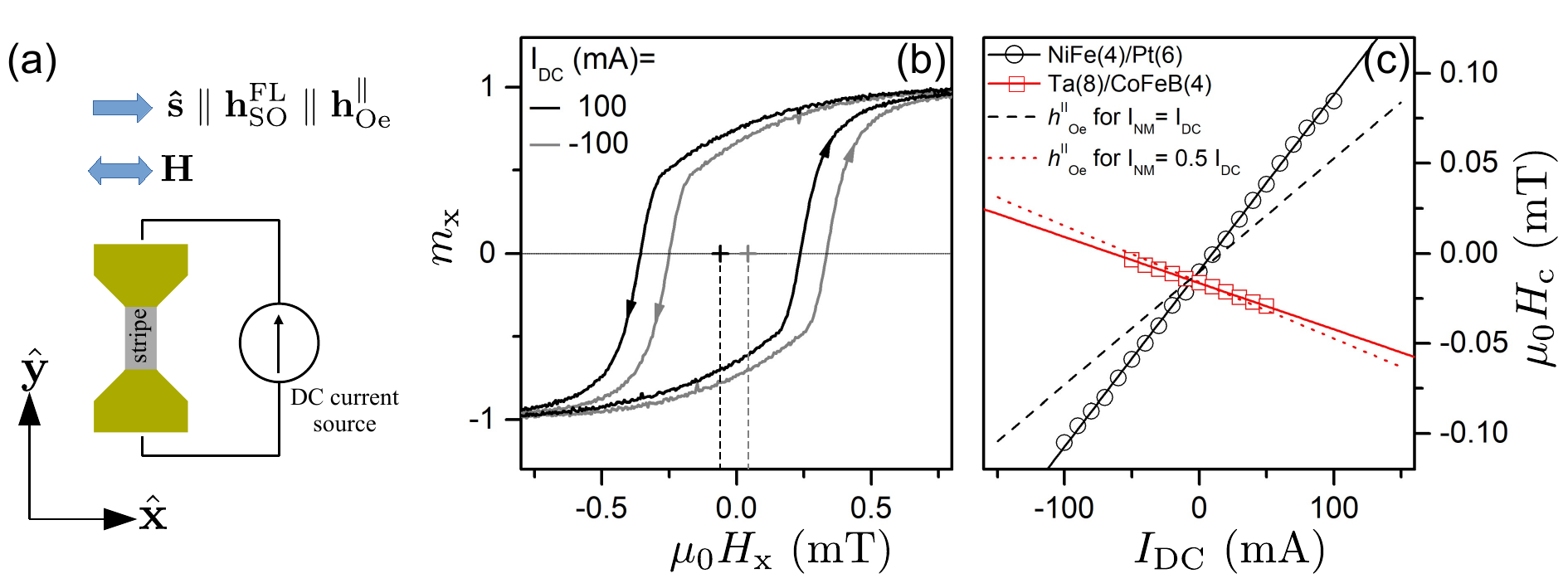}
\caption{\raggedright (a)  Schematic of the sample arrangement employed for $\mathbf{h}^{\text{DL}}_{\text{SO}}$ measurement. (b) Longitudinal MOKE H-loops of the NiFe4 stripe, upon the application of continuous electrical current: $I_{\text{DC}}=100$ mA (black curve) and $I_{\text{DC}}=-100$ mA (gray curve). The crosses indicate the center of symmetry of each cycle, while the dashed lines indicate the corresponding values for $\mu_{0}H_{\text{c}}$ on the horizontal axis. (c) $\mu_{0}H_{\text{c}}$ vs $I_{\text{DC}}$ for  NiFe(4)/Pt(6) and Ta(8)/CoFeB(4) samples. Open symbols denote experimental points, and continuous lines correspond to the linear fittings. Discontinuous lines are the expected curves when only the Oersted field is acting and all the electrical current passes through the NM layer (black dashed line), or half of the current passes through the NM layer (dotted red line).}
\label{fig3}
\end{figure*}
\end{center}

For FL-SOT measurement, the device is set to work as a conventional MOKE hysteresis-loop tracer in longitudinal configuration, using
the sample and field geometry depicted in (Fig. \ref{fig3}{\color{blue}{(a)}}). We set the polarizer angles to a fixed value $(\varphi_{1}=0^{\circ},\varphi_{2}=0.75^{\circ}$), which has a convenient signal-to-noise ratio for MOKE. In this configuration, the variable component of the photo-voltage is proportional to the $x$-axis component of the normalized magnetization vector ($m_{x}$) \cite{Gonzalez-Fuentes20}.
Helmholtz coils are fed with a low-frequency electrical current (16 Hz), previously calibrated to deliver external field $\mathbf{H}=H_{x}\hat{\mathbf{x}}$ of known value in the sample's position.
According to the system of reference depicted in Fig. \ref{fig3}{\color{blue}{(a)}}, spin current diffusing from the NM to the FM layer has a spin polarization vector ($\mathbf{\hat{s}}$) is parallel to $\mathbf{\hat{x}}$. The total FL effective field sensed by the FM layer will be given by:
\begin{equation}
\mathbf{h}^{\text{FL}}=\left(h^{\text{FL}}_{\text{SO}}+h^{\parallel}_{\text{Oe}}\right)\mathbf{\hat{x}}.
\label{hFL}
\end{equation}
Here $h^{\text{FL}}_{\text{SO}}$ and $h^{\parallel}_{\text{Oe}}$ are scalars that parametrize the in-plane component of $\mathbf{h}^{\text{FL}}_{\text{SO}}$ and the Oersted field, respectively.

As shown in Fig. \ref{fig3},{\color{blue}{(a)}}, $\mathbf{h}^{\text{FL}}$ induces a homogeneous $H$-axis shifting on the MOKE H-loop curve under a DC current\cite{Xing20}.
To quantify this shifting, we choose $H_{\text{c}}$ as a reference, defined as the value of the $H$-axis where the center of symmetry of each $H$-loop is located. We take it as $H_{\text{c}}=(H_{\uparrow}+H_{\downarrow})/2$, where $H_{\uparrow}$ ($H_{\downarrow}$) is the $H$-axis value where $m_{x}$ crosses from negative (positive) to positive (negative) region of the cycle.  Fig. \ref{fig3}{\color{blue}{(c)}} shows  $H_c$ vs $I_{\text{DC}}$ cures for NiFe4 and CoFeB4 stripes, showing excellent linearity. 

For the analysis, we assume $h^{\text{FL}}/I_{\text{DC}}=dH_{\text{c}}/d I_{\text{DC}}$, in this way, we eliminate background DC magnetic field contributions that may shift $H_{\text{c}}$. Using Eq. \eqref{hFL}, we  can establish the following relation:
\begin{equation}
h^{\text{FL}}_{\text{SO}}=I_{\text{DC}}\left(\frac{dH_{\text{c}}}{dI_{\text{DC}}}-\frac{I_{\text{NM}}}{2I_{\text{DC}}w}\right),
\label{fl}
\end{equation}
where $I_{\text{NM}}/I_{\text{DC}}$ ratio was obtained from the planar resistivities of and thicknesses of FM and NN layers \cite{Boone15}  (see supplementary material).

The effect of $\mathbf{h}^{\text{FL}}_{\text{SO}}$ on the thinner group of ferromagnetic films can be directly inferred from $H_{\text{c}}$ vs $I_{\text{DC}}$ curves plotted on  Fig. \ref{fig3}{\color{blue}{(c)}}. We see that in the NiFePt(4)/Pt(6) sample, $dH_{\text{c}}/dI_{\text{DC}}$ surpasses by 62\% the maximum value attributable to $\mathbf{h}^{\parallel}_{\text{Oe}}$, in the case in which and all the electrical current were passing through the NM layer i.e. $I_{\text{NM}}/I_{\text{DC}}$=1 (black, dashed line). On the other side, for the Ta(8)/CoFeB4(4) sample, $\mathbf{h}^{\parallel}_{\text{Oe}}$ and $\mathbf{h}^{\text{FL}}_{\text{SO}}$ have opposite signs, leading to partial cancellation of the first, resulting in $\mu_{0}dH_{\text{c}}/dI_{\text{DC}}=\text{-0.23}\text{ mT/A}$. This value is significantly below the attributable to $\mathbf{h}^{\parallel}_{\text{Oe}}$ with a current distribution of $I_{\text{NM}}/I_{\text{DC}}=0.5$ (red, dotted line), which is the expected ratio according to the relative resistivities and thicknesses of Ta and CoFeB layers. Note also that in this case, the sign of $dH_{\text{c}}/dI_{\text{DC}}$ is negative, given that the NM layer is at the bottom of the bilayer. The emergence of  $\mathbf{h}^{\text{FL}}$ in ferromagnetic films of similar thickness has also been reported previously\cite{Fan13,Emori16}.

\begin{center}
 \begin{figure*}[ht]
\includegraphics[width=14cm]{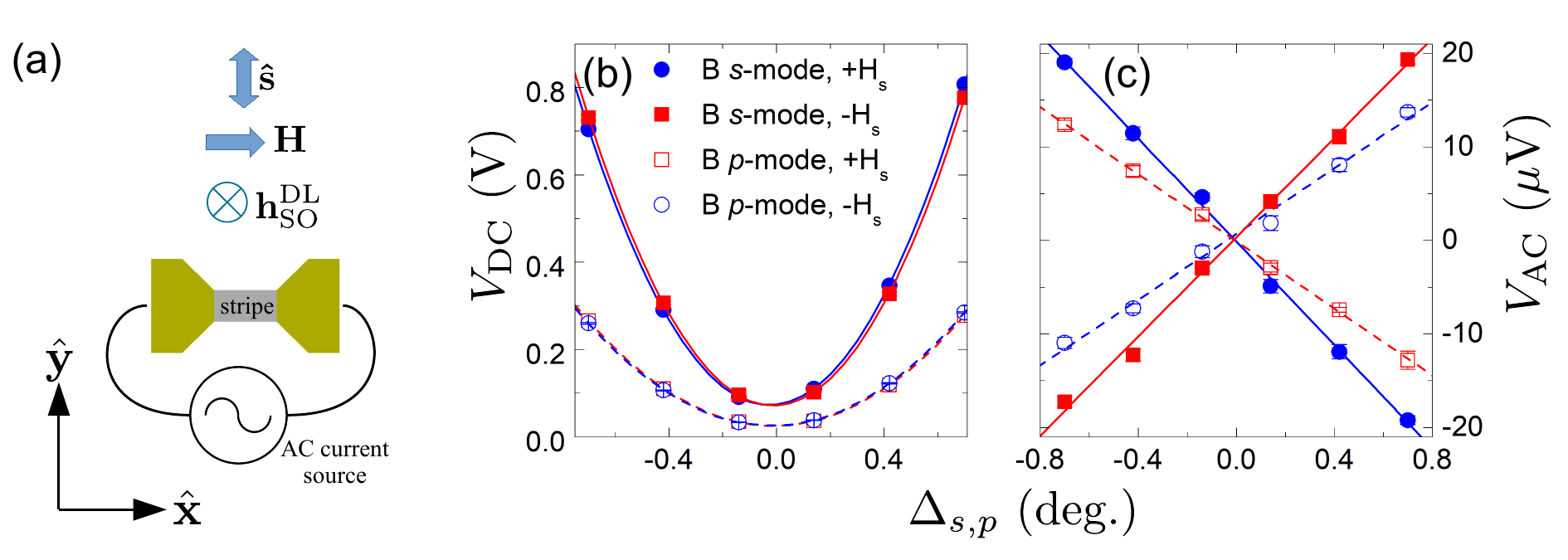}
\caption{\raggedright (a) Schematic of the sample arrangement employed for $\mathbf{h}^{\text{DL}}_{\text{SO}}$ measurement. (b-c) DC and AC components, respectively, of the photo-voltage induced by the reflected light from the sample configured like in (a). The magnetization is saturated along $\pm \mathbf{x}$ direction by an external field of 14 mT magnitude.
}
\label{fig2}
\end{figure*}
\end{center}

Now, we describe the DL-SOT measurement. In this case, we employ the geometry depicted in Fig. \ref{fig2}{\color{blue}{(a)}}
. The external magnetic field is set to a fixed value $\mathbf{H}=H_{\text{s}}\mathbf{\hat{x}}$, with $H_{\text{s}}=14$ mT, which is sufficient to saturate the magnetization parallel to $x$-axis. Simultaneously, an oscillating current (824 Hz) is injected into the stripe, generating a total SOT effective field given by: $\left( h^{\text{DL}}_{\text{SO}}\mathbf{\hat{z}}+h^{\text{FL}}_{\text{SO}}\bf{\hat{y}}\right)\cos\left(\omega t\right)$.
The z-axis periodic deviations induce a polar Kerr rotation of the form $\theta\cos\left(\omega t\right)$, which can be straightforwardly detected via lock-in amplification. However, the concurrent $y$-axis deviations due to $h^{\text{FL}}_{\text{SO}}$ component in addition to the Oersted field for this geometry: $\mathbf{h}_{\text{Oe}}=\left(h^{\perp}_{\text{Oe}}\mathbf{\hat{z}}+h^{\parallel}_{\text{Oe}}\bf{\hat{y}}\right)\cos\left(\omega t\right)$, will produce not only polar but also and longitudinal-transversal MOKE signal synchronously, mixing-up in the lock-in detected in-phase term.

In this work, we employ a novel approach to isolate the $\theta$ component exclusively from $\mathbf{h}^{\text{DL}}_{\text{SO}}$. We execute the measurements in two stages: in the first (second) stage, we keep the P1 (P2) axis fixed parallel to the s(p)-polarization axis while the P2 (P1) axis is varied slightly around the s(p)-polarization axis by an angle $\Delta_{s}(\Delta_{p})$. In this manner, we tune the amount of $s(p)$-polarized light that passes through the second(first) polarizer. We denote the first and second measurement modes as the $s$-mode and $p$-mode, respectively.  
Both $\Delta_{s}$ and $\Delta_{p}$ were in a range of less than 0.7 degrees, thus very close to the extinction condition of light. 

In both measurement modes, we will have a photo-detected voltage of the form: $ V_{\text{D}}=V_{\text{DC}}+V_{\text{AC}}\cos\left(\omega t\right)$. The DC component ($V_{\text{DC}}$) is related to the reflectometry of the sample. In contrast, the AC component ($V_{\text{AC}}$) is linked to $\theta$ owing to deviations in the magnetization direction from equilibrium.  Fig. \ref{fig2}{\color{blue}{(b)}} ({\color{blue}{(c)}}) shows experimental $V_{\text{DC}}$ ($V_{\text{AC}}$) curve, where we observe a dominant quadratic (linear) dependence on $\Delta_{s(p)}$).
From these curves and Jones matrix analysis (see supplemental material), we extract $h_{\text{SO}}^{\text{DL}}$ by the following expression:
\begin{align}
\theta=\alpha^{\text{eff}}_{\text{P}}h^{\text{DL}}_{\text{SO}}=
&\frac{1}{4}\operatorname{Re}\left[\theta_{s}^{+H} - \theta_{s}^{-H}+\theta_{p}^{+H}-\theta_{p}^{-H}\right], \label{aeff}
\end{align}
with:
\begin{equation}
\theta_{s(p)}=\left.\pm\left(\frac{\partial V_{\text{AC}}}{\partial \Delta_{s(p)}}\right)\right/\left(\frac{\partial^{2} V_{\text{DC}}}{\partial \Delta_{s(p)}^{2}}\right) \label{rotsp}
\end{equation}
where subindexes denote measurements done in {\it{s}}({\it{p}})-mode and super-indexes $+H$($-H$) indicates measurements done with $H_{\text{s}}>0$ ($H_{\text{s}}<0$) i.e. with opposite directions of saturation magnetization. 
The obtention of $h_{\text{SO}}^{\text{DL}}$ in a single sample thus requires the analysis of 8 experimental curves: $V_{\text{AC}}$ and $V_{\text{DC}}$ for each term of the right side of Eq. \ref{aeff}. 
In addition, $\alpha_{\text{eff}}$ is a sample-specific calibration coefficient that parametrizes the ratio between polar MOKE and out-of-plane magnetic field excitations. It must be obtained independently by applying an AC magnetic field of known amplitude $h_{z}$= 4 mT that replaces $h_{\text{SO}}^{\text{DL}}$ in Eq. \ref{aeff}. A representative example of the experimental and fitting curves is shown in Fig. \ref{fig2}, with the NiFe(4)/Pt(6) sample.

We note that our measurement protocol, summarized in the terms of Eqs. \ref{aeff} is designed to thoroughly eliminate prop-to $\cos\omega t$ signals that may distort the accurate quantification of $\mathbf{h}_{\text{SO}}^{\text{DL}}$. 
The {\it{s}}-mode {\it{p}}-mode averaging eliminates the second-order or quadratic MOKE, which may be comparable to polar MOKE for in-plane magnetized samples \cite{Fan16}. In addition, the field reversal averaging cancels out $\mathbf{h}_{\text{Oe}}^{\perp}$, given that is sign does not reverse with $\mathbf{m}$ reversal, but $\mathbf{h}^{\text{DL}}_{\text{SO}}$ does. Compared to previous works that employ normal incidence angle of light \cite{Fan16}, no modifications in the setup are required to move from the DL-SOT to the FL-SOT measurement mode. The changes are only in the electrical feed-through of the stripe, the Helmholtz coils, and the activation/deactivation of the automated rotation of polarizer axes.  All the steps, except the change of samples, can be accomplished without hardware modifications so that it can be automatized with appropriate programming control. 

Continuing with the analysis of DL-SOT, we assumed that in all the samples, the NM spin diffusion length is significantly shorter than $t_{\text{NM}}$ \cite{Gonzalez-Fuentes21,Foros05,Allen15}. We then can employ the following approximation for damping-like ($\xi^{\text{DL}}$) and field-like ($\xi^{\text{FL}}$) effective spin-orbit torque efficiencies \cite{Nguyen21}: 
\begin{equation}
\xi^{\text{DL(FL)}}= \mu_{0}M_{\text{s}}t_{\text{FM}}\left(\frac{2q_{e}}{\hbar}\right)\frac{h^{\text{DL(FL)}}_{\text{SO}}}{J_{\text{NM}}},
\label{xi}
\end{equation}
where $M_{s}$ is the saturation magnetization of the FM layer. We employed $\mu_{0}M_{s}=$  0.98 T and 1.28 for NiFe and CoFeB, respectively. These values were extracted from FMR measurements performed on bulk samples fabricated in the same system and conditions.
In Table \ref{results}, we list our extracted values of $\xi^{\text{DL}}$ for each system. Overall, we find a good agreement of $\xi^{\text{DL}}$ to the reported values for bilayers with the same combination FM/NM layers: 0.06 to 0.15 for NiFe/Pt \cite{Soya21, Shashank21,Karimeddiny21}, -0.18 to -0.11 for CoFeB/Ta \cite{Liu12,Allen15,Huang18}. In the case of Pd, $\xi^{\text{DL}}$ is roughly a fourth of the Pt value, which is consistent with previous reports and the relative strength of SOC of Pd with respect to Pt \cite{Kondou12,Tao18}. 
 We also did not see significant variations on $\xi^{\text{DL}}$ between the NiFe4 and NiFe8 samples nor between CoFeB4 and CoFeB8 samples.

Regarding the field-like component of SOT, we found that in NM=Pt samples $\mathbf{h}^{\text{FL}}_\text{SO}$ has the same sign as $\mathbf{h}^{\parallel}_\text{Oe}$, whereas for NM=Ta is the opposite, in line with previous observations \cite{Fan13,Fan14,Allen15,Liu12}. 


Table \ref{results} shows a comparison of the relative strengths of the FL and DL SOTs in our samples, parametrized by $\xi^{\text{FL}}/\xi^{\text{DL}}$ value,  as well as a comparison with previous reports on NM/FM bilayers with identical composition. The evaluation of this parameter instead of the absolute value of $\xi^{\text{DL(FL)}}$ is very relevant given that the latter can be affected by the spin memory loss of the interface \cite{Rojas-Sanchez14} or the resistivity of the NM layer \cite{Sagasta16, Nguyen16}, so it can be notably affected by the variability of fabrication conditions. We also note that the high resistivity of the Ta 8nm layer (243 $\mu \Omega\text{m}$) lets us assume that it grew in the $\beta$ phase (see supplemental material).

Overall, Table \ref{results}, shows that $\xi^{\text{FL}}/\xi^{\text{DL}}$ is a strictly decreasing function on $t_{\text{FM}}$ and our results fits well inside this trend. This is in agreement with the classic diffusive model of spin current absorption in an FM layer \cite{Zhang02} with a finite spin dephasing length $\lambda_{\text{J}}$: the characteristic length over which the spins diffusing into FM layer align with $\mathbf{m}$ owing to the exchange interaction. This phenomenon can be a significant source of FL-SOT in FM layers of thickness comparable to $\lambda_{\text{J}}$, which is of the order of few nm \cite{Zhang02,Manchon12} for typical 3d-group ferromagnets. In consequence,  $h_{\text{SO}}^{\text{FL}}$ vanishes asymptotically as $t_{\text{FM}}\gg \lambda_{\text{J}}$, whereas $h_{\text{SO}}^{\text{DL}}$ follows the natural $1/t_{\text{NM}}$ dependence. In consequence $\xi^{\text{FL}}/\xi^{\text{DL}}$ is a strictly and rapidly deceasing function of $t_{\text{FM}}$.
\begin{center}
\begin{table}[t!]
\begin{tabular}{>{\centering\arraybackslash}m{1.8cm} >{\centering\arraybackslash}m{1.3cm} >{\centering\arraybackslash}m{1.6cm}>{\centering\arraybackslash}m{1.8cm} >{\centering\arraybackslash}m{1.8cm} }
\hline
\hline \\[-1.1em]
Layer structure & \multicolumn{1}{|c}{$t_{\text{FM}}$} & $\xi^{\text{DL}}$ &  $\xi^{\text{FL}}/\xi^{\text{DL}}$  &  \multicolumn{1}{|c}{Ref.} \\ \\[-1.2em] 
\multicolumn{1}{c}{}  & \multicolumn{1}{|c}{(nm)}      
 & (\%)  &   &  \multicolumn{1}{|c}{} \\ \cline{1-5} 
\\[0.1em]
NiFe/Pt    & 2.         &    7                           & 0.42                        &     \cite{Fan13} \\
              & 2.5      & 8.7 $\pm$ 0.7          & 0.27                         &   \cite{Nan15} \\ 
              & 4        & 8.2 $\pm$ 0.3          & 0.35 $\pm$ 0.02       & This work.  \\
              & 8        & 8.9 $\pm$ 0.6          &  -                            &  This work \\
              &  6-9    &  5. $\pm$ 0.5            & 0.07                       &    \cite{Haku20} \\
              &          &                                &                                & \\ \hline \\[-0.4em]
NiFe/Pd & 8          & 1.9 $\pm$ 0.2            &  -  &   This work. \\
               & 10          & 1.0     &  -  &   \cite{Ando10} \\
              &             &                              &                              &\\ \hline \\[-0.4em]
$\beta$-Ta/CoFeB    & 0.75-1.3                    &                           &                 1-10             &   \cite{Kim13}\\
              & 4          & -11 $\pm$ 1               &  0.16       $\pm$ 0.03              &   \cite{Allen15} \\ 
              & 4          &  -12.7  $\pm$ 0.8         &   0.14 $\pm$ 0.02 &  This work.  \\              
              & 8          &  -13.2  $\pm$ 0.9        &   -                      &  This work.  \\           
 \hline \hline
\end{tabular}
\caption{Comparision of $\xi^{\text{DL}}$, $\xi^{\text{FL}}$ and $\xi^{\text{FL}}/\xi^{\text{DL}}$ values obtained in our samples vs the previously reported on bilayers of equal composition.}
\label{results}
\end{table}
\end{center}
In summary, we have demonstrated the magneto-optical detection of field-like and damping-like spin-orbit torque vectors in different combinations of normal-metal/ferromagnetic-metal bilayers
with a simple and cost-effective setup. 

Notably, our methodology relies only on first-order magneto-optical Kerr effect coefficients, longitudinal and polar, which, unlike second-order ones, scale directly with the material's magnetization, ensuring broad applicability to any type of ferromagnetic material.

We obtained damping-like spin-orbit torque efficiency $\xi_{\text{DL}}=0.089 \pm 0.006$, $0.019 \pm 0.02$, and $-0.132 \pm 0.009$ for Pt, Pd, and Ta, respectively. These values are in good agreement, in sign and magnitude, with the most accepted values for these metals.

The ratio between the field-like and damping-like spin-orbit torque efficiency $\xi^{\text{FL}}/\xi^{\text{DL}}$, was $0.35 \pm 0.02$  and $0.14 \pm 0.02$ for NiFe/Pt and CoFeB/Ta bilayers, respectively, when the thickness of the ferromagnetic layer is 4 nm. 

We anticipate this work will benefit the broader adoption of magneto-optical probing of SOTs, given its intrinsic advantages over electrical detection methods, such as its robust isolation of magnetization vector components and its imperturbability by thermoelectric artifacts.

\section*{Acknowledgments}
This work has been supported by FONDECYT 11220854, ANID PIA/APOYO AFB220003, ANID SIA 85220125, and FONDEQUIP EQM140161.
\bibliographystyle{apsrev4-2}
\bibliography{bibliography}

\begin{thebibliography}{60}%
\makeatletter
\providecommand \@ifxundefined [1]{%
 \@ifx{#1\undefined}
}%
\providecommand \@ifnum [1]{%
 \ifnum #1\expandafter \@firstoftwo
 \else \expandafter \@secondoftwo
 \fi
}%
\providecommand \@ifx [1]{%
 \ifx #1\expandafter \@firstoftwo
 \else \expandafter \@secondoftwo
 \fi
}%
\providecommand \natexlab [1]{#1}%
\providecommand \enquote  [1]{``#1''}%
\providecommand \bibnamefont  [1]{#1}%
\providecommand \bibfnamefont [1]{#1}%
\providecommand \citenamefont [1]{#1}%
\providecommand \href@noop [0]{\@secondoftwo}%
\providecommand \href [0]{\begingroup \@sanitize@url \@href}%
\providecommand \@href[1]{\@@startlink{#1}\@@href}%
\providecommand \@@href[1]{\endgroup#1\@@endlink}%
\providecommand \@sanitize@url [0]{\catcode `\\12\catcode `\$12\catcode
  `\&12\catcode `\#12\catcode `\^12\catcode `\_12\catcode `\%12\relax}%
\providecommand \@@startlink[1]{}%
\providecommand \@@endlink[0]{}%
\providecommand \url  [0]{\begingroup\@sanitize@url \@url }%
\providecommand \@url [1]{\endgroup\@href {#1}{\urlprefix }}%
\providecommand \urlprefix  [0]{URL }%
\providecommand \Eprint [0]{\href }%
\providecommand \doibase [0]{https://doi.org/}%
\providecommand \selectlanguage [0]{\@gobble}%
\providecommand \bibinfo  [0]{\@secondoftwo}%
\providecommand \bibfield  [0]{\@secondoftwo}%
\providecommand \translation [1]{[#1]}%
\providecommand \BibitemOpen [0]{}%
\providecommand \bibitemStop [0]{}%
\providecommand \bibitemNoStop [0]{.\EOS\space}%
\providecommand \EOS [0]{\spacefactor3000\relax}%
\providecommand \BibitemShut  [1]{\csname bibitem#1\endcsname}%
\let\auto@bib@innerbib\@empty
\bibitem [{\citenamefont {Dyakonov}\ and\ \citenamefont
  {Perel}(1971)}]{Dyakonov71}%
  \BibitemOpen
  \bibfield  {author} {\bibinfo {author} {\bibfnamefont {M.~I.}\ \bibnamefont
  {Dyakonov}}\ and\ \bibinfo {author} {\bibfnamefont {V.~I.}\ \bibnamefont
  {Perel}},\ }\href {http://www.jetpletters.ac.ru/ps/1587/article_24366.shtml}
  {\bibfield  {journal} {\bibinfo  {journal} {Sov. Phys. JETP Lett.}\ }\textbf
  {\bibinfo {volume} {13}},\ \bibinfo {pages} {467} (\bibinfo {year}
  {1971})}\BibitemShut {NoStop}%
\bibitem [{\citenamefont {Kato}\ \emph {et~al.}(2004)\citenamefont {Kato},
  \citenamefont {Myers}, \citenamefont {Gossard},\ and\ \citenamefont
  {Awschalom}}]{Kato04}%
  \BibitemOpen
  \bibfield  {author} {\bibinfo {author} {\bibfnamefont {Y.~K.}\ \bibnamefont
  {Kato}}, \bibinfo {author} {\bibfnamefont {R.~C.}\ \bibnamefont {Myers}},
  \bibinfo {author} {\bibfnamefont {A.~C.}\ \bibnamefont {Gossard}},\ and\
  \bibinfo {author} {\bibfnamefont {D.~D.}\ \bibnamefont {Awschalom}},\ }\href
  {https://doi.org/10.1126/science.1105514} {\bibfield  {journal} {\bibinfo
  {journal} {Science}\ }\textbf {\bibinfo {volume} {306}},\ \bibinfo {pages}
  {1910} (\bibinfo {year} {2004})},\ \Eprint
  {https://arxiv.org/abs/https://www.science.org/doi/pdf/10.1126/science.1105514}
  {https://www.science.org/doi/pdf/10.1126/science.1105514} \BibitemShut
  {NoStop}%
\bibitem [{\citenamefont {Manchon}\ \emph {et~al.}(2019)\citenamefont
  {Manchon}, \citenamefont {\ifmmode~\check{Z}\else \v{Z}\fi{}elezn\'y},
  \citenamefont {Miron}, \citenamefont {Jungwirth}, \citenamefont {Sinova},
  \citenamefont {Thiaville}, \citenamefont {Garello},\ and\ \citenamefont
  {Gambardella}}]{Manchon19}%
  \BibitemOpen
  \bibfield  {author} {\bibinfo {author} {\bibfnamefont {A.}~\bibnamefont
  {Manchon}}, \bibinfo {author} {\bibfnamefont {J.}~\bibnamefont
  {\ifmmode~\check{Z}\else \v{Z}\fi{}elezn\'y}}, \bibinfo {author}
  {\bibfnamefont {I.~M.}\ \bibnamefont {Miron}}, \bibinfo {author}
  {\bibfnamefont {T.}~\bibnamefont {Jungwirth}}, \bibinfo {author}
  {\bibfnamefont {J.}~\bibnamefont {Sinova}}, \bibinfo {author} {\bibfnamefont
  {A.}~\bibnamefont {Thiaville}}, \bibinfo {author} {\bibfnamefont
  {K.}~\bibnamefont {Garello}},\ and\ \bibinfo {author} {\bibfnamefont
  {P.}~\bibnamefont {Gambardella}},\ }\href
  {https://doi.org/10.1103/RevModPhys.91.035004} {\bibfield  {journal}
  {\bibinfo  {journal} {Rev. Mod. Phys.}\ }\textbf {\bibinfo {volume} {91}},\
  \bibinfo {pages} {035004} (\bibinfo {year} {2019})}\BibitemShut {NoStop}%
\bibitem [{\citenamefont {Bhatti}\ \emph {et~al.}(2017)\citenamefont {Bhatti},
  \citenamefont {Sbiaa}, \citenamefont {Hirohata}, \citenamefont {Ohno},
  \citenamefont {Fukami},\ and\ \citenamefont {Piramanayagam}}]{Bhatti17}%
  \BibitemOpen
  \bibfield  {author} {\bibinfo {author} {\bibfnamefont {S.}~\bibnamefont
  {Bhatti}}, \bibinfo {author} {\bibfnamefont {R.}~\bibnamefont {Sbiaa}},
  \bibinfo {author} {\bibfnamefont {A.}~\bibnamefont {Hirohata}}, \bibinfo
  {author} {\bibfnamefont {H.}~\bibnamefont {Ohno}}, \bibinfo {author}
  {\bibfnamefont {S.}~\bibnamefont {Fukami}},\ and\ \bibinfo {author}
  {\bibfnamefont {S.}~\bibnamefont {Piramanayagam}},\ }\href
  {https://doi.org/https://doi.org/10.1016/j.mattod.2017.07.007} {\bibfield
  {journal} {\bibinfo  {journal} {Materials Today}\ }\textbf {\bibinfo {volume}
  {20}},\ \bibinfo {pages} {530} (\bibinfo {year} {2017})}\BibitemShut
  {NoStop}%
\bibitem [{\citenamefont {Torrejon}\ \emph {et~al.}(2017)\citenamefont
  {Torrejon}, \citenamefont {Riou}, \citenamefont {Araujo}, \citenamefont
  {Tsunegi}, \citenamefont {Khalsa}, \citenamefont {Querlioz}, \citenamefont
  {Bortolotti}, \citenamefont {Cros}, \citenamefont {Yakushiji}, \citenamefont
  {Fukushima}, \citenamefont {Kubota}, \citenamefont {Yuasa}, \citenamefont
  {Stiles},\ and\ \citenamefont {Grollier}}]{Torrejon17}%
  \BibitemOpen
  \bibfield  {author} {\bibinfo {author} {\bibfnamefont {J.}~\bibnamefont
  {Torrejon}}, \bibinfo {author} {\bibfnamefont {M.}~\bibnamefont {Riou}},
  \bibinfo {author} {\bibfnamefont {F.~A.}\ \bibnamefont {Araujo}}, \bibinfo
  {author} {\bibfnamefont {S.}~\bibnamefont {Tsunegi}}, \bibinfo {author}
  {\bibfnamefont {G.}~\bibnamefont {Khalsa}}, \bibinfo {author} {\bibfnamefont
  {D.}~\bibnamefont {Querlioz}}, \bibinfo {author} {\bibfnamefont
  {P.}~\bibnamefont {Bortolotti}}, \bibinfo {author} {\bibfnamefont
  {V.}~\bibnamefont {Cros}}, \bibinfo {author} {\bibfnamefont {K.}~\bibnamefont
  {Yakushiji}}, \bibinfo {author} {\bibfnamefont {A.}~\bibnamefont
  {Fukushima}}, \bibinfo {author} {\bibfnamefont {H.}~\bibnamefont {Kubota}},
  \bibinfo {author} {\bibfnamefont {S.}~\bibnamefont {Yuasa}}, \bibinfo
  {author} {\bibfnamefont {M.~D.}\ \bibnamefont {Stiles}},\ and\ \bibinfo
  {author} {\bibfnamefont {J.}~\bibnamefont {Grollier}},\ }\href
  {https://doi.org/10.1038/nature23011} {\bibfield  {journal} {\bibinfo
  {journal} {Nature}\ }\textbf {\bibinfo {volume} {547}},\ \bibinfo {pages}
  {428} (\bibinfo {year} {2017})}\BibitemShut {NoStop}%
\bibitem [{\citenamefont {Saitoh}\ \emph {et~al.}(2006)\citenamefont {Saitoh},
  \citenamefont {Ueda}, \citenamefont {Miyajima},\ and\ \citenamefont
  {Tatara}}]{Saitoh06}%
  \BibitemOpen
  \bibfield  {author} {\bibinfo {author} {\bibfnamefont {E.}~\bibnamefont
  {Saitoh}}, \bibinfo {author} {\bibfnamefont {M.}~\bibnamefont {Ueda}},
  \bibinfo {author} {\bibfnamefont {H.}~\bibnamefont {Miyajima}},\ and\
  \bibinfo {author} {\bibfnamefont {G.}~\bibnamefont {Tatara}},\ }\href
  {https://doi.org/10.1063/1.2199473} {\bibfield  {journal} {\bibinfo
  {journal} {Applied Physics Letters}\ }\textbf {\bibinfo {volume} {88}},\
  \bibinfo {pages} {182509} (\bibinfo {year} {2006})}\BibitemShut {NoStop}%
\bibitem [{\citenamefont {Pi}\ \emph {et~al.}(2010)\citenamefont {Pi},
  \citenamefont {Won~Kim}, \citenamefont {Bae}, \citenamefont {Lee},
  \citenamefont {Cho}, \citenamefont {Kim},\ and\ \citenamefont {Seo}}]{Pi10}%
  \BibitemOpen
  \bibfield  {author} {\bibinfo {author} {\bibfnamefont {U.~H.}\ \bibnamefont
  {Pi}}, \bibinfo {author} {\bibfnamefont {K.}~\bibnamefont {Won~Kim}},
  \bibinfo {author} {\bibfnamefont {J.~Y.}\ \bibnamefont {Bae}}, \bibinfo
  {author} {\bibfnamefont {S.~C.}\ \bibnamefont {Lee}}, \bibinfo {author}
  {\bibfnamefont {Y.~J.}\ \bibnamefont {Cho}}, \bibinfo {author} {\bibfnamefont
  {K.~S.}\ \bibnamefont {Kim}},\ and\ \bibinfo {author} {\bibfnamefont
  {S.}~\bibnamefont {Seo}},\ }\href {https://doi.org/10.1063/1.3502596}
  {\bibfield  {journal} {\bibinfo  {journal} {Applied Physics Letters}\
  }\textbf {\bibinfo {volume} {97}},\ \bibinfo {pages} {162507} (\bibinfo
  {year} {2010})},\ \Eprint
  {https://arxiv.org/abs/https://doi.org/10.1063/1.3502596}
  {https://doi.org/10.1063/1.3502596} \BibitemShut {NoStop}%
\bibitem [{\citenamefont {Amin}\ \emph {et~al.}(2018)\citenamefont {Amin},
  \citenamefont {Zemen},\ and\ \citenamefont {Stiles}}]{Amin18}%
  \BibitemOpen
  \bibfield  {author} {\bibinfo {author} {\bibfnamefont {V.~P.}\ \bibnamefont
  {Amin}}, \bibinfo {author} {\bibfnamefont {J.}~\bibnamefont {Zemen}},\ and\
  \bibinfo {author} {\bibfnamefont {M.~D.}\ \bibnamefont {Stiles}},\ }\href
  {https://doi.org/10.1103/PhysRevLett.121.136805} {\bibfield  {journal}
  {\bibinfo  {journal} {Phys. Rev. Lett.}\ }\textbf {\bibinfo {volume} {121}},\
  \bibinfo {pages} {136805} (\bibinfo {year} {2018})}\BibitemShut {NoStop}%
\bibitem [{\citenamefont {Baek}\ \emph {et~al.}(2018)\citenamefont {Baek},
  \citenamefont {Amin}, \citenamefont {Oh}, \citenamefont {Go}, \citenamefont
  {Lee}, \citenamefont {Lee}, \citenamefont {Kim}, \citenamefont {Stiles},
  \citenamefont {Park},\ and\ \citenamefont {Lee}}]{Baek18}%
  \BibitemOpen
  \bibfield  {author} {\bibinfo {author} {\bibfnamefont {S.-h.~C.}\
  \bibnamefont {Baek}}, \bibinfo {author} {\bibfnamefont {V.~P.}\ \bibnamefont
  {Amin}}, \bibinfo {author} {\bibfnamefont {Y.-W.}\ \bibnamefont {Oh}},
  \bibinfo {author} {\bibfnamefont {G.}~\bibnamefont {Go}}, \bibinfo {author}
  {\bibfnamefont {S.-J.}\ \bibnamefont {Lee}}, \bibinfo {author} {\bibfnamefont
  {G.-H.}\ \bibnamefont {Lee}}, \bibinfo {author} {\bibfnamefont {K.-J.}\
  \bibnamefont {Kim}}, \bibinfo {author} {\bibfnamefont {M.~D.}\ \bibnamefont
  {Stiles}}, \bibinfo {author} {\bibfnamefont {B.-G.}\ \bibnamefont {Park}},\
  and\ \bibinfo {author} {\bibfnamefont {K.-J.}\ \bibnamefont {Lee}},\ }\href
  {https://doi.org/10.1038/s41563-018-0041-5} {\bibfield  {journal} {\bibinfo
  {journal} {Nature Materials}\ }\textbf {\bibinfo {volume} {17}},\ \bibinfo
  {pages} {509} (\bibinfo {year} {2018})}\BibitemShut {NoStop}%
\bibitem [{\citenamefont {Luo}\ \emph {et~al.}(2019)\citenamefont {Luo},
  \citenamefont {Zhang}, \citenamefont {Xu}, \citenamefont {Yang},
  \citenamefont {Zhang},\ and\ \citenamefont {Wu}}]{Luo19}%
  \BibitemOpen
  \bibfield  {author} {\bibinfo {author} {\bibfnamefont {Z.}~\bibnamefont
  {Luo}}, \bibinfo {author} {\bibfnamefont {Q.}~\bibnamefont {Zhang}}, \bibinfo
  {author} {\bibfnamefont {Y.}~\bibnamefont {Xu}}, \bibinfo {author}
  {\bibfnamefont {Y.}~\bibnamefont {Yang}}, \bibinfo {author} {\bibfnamefont
  {X.}~\bibnamefont {Zhang}},\ and\ \bibinfo {author} {\bibfnamefont
  {Y.}~\bibnamefont {Wu}},\ }\href
  {https://doi.org/10.1103/PhysRevApplied.11.064021} {\bibfield  {journal}
  {\bibinfo  {journal} {Phys. Rev. Applied}\ }\textbf {\bibinfo {volume}
  {11}},\ \bibinfo {pages} {064021} (\bibinfo {year} {2019})}\BibitemShut
  {NoStop}%
\bibitem [{\citenamefont {Sala}\ and\ \citenamefont
  {Gambardella}(2022)}]{Sala22}%
  \BibitemOpen
  \bibfield  {author} {\bibinfo {author} {\bibfnamefont {G.}~\bibnamefont
  {Sala}}\ and\ \bibinfo {author} {\bibfnamefont {P.}~\bibnamefont
  {Gambardella}},\ }\href {https://doi.org/10.1103/PhysRevResearch.4.033037}
  {\bibfield  {journal} {\bibinfo  {journal} {Phys. Rev. Res.}\ }\textbf
  {\bibinfo {volume} {4}},\ \bibinfo {pages} {033037} (\bibinfo {year}
  {2022})}\BibitemShut {NoStop}%
\bibitem [{\citenamefont {Kim}\ \emph {et~al.}(2013)\citenamefont {Kim},
  \citenamefont {Sinha}, \citenamefont {Hayashi}, \citenamefont {Yamanouchi},
  \citenamefont {Fukami}, \citenamefont {Suzuki}, \citenamefont {Mitani},\ and\
  \citenamefont {Ohno}}]{Kim13}%
  \BibitemOpen
  \bibfield  {author} {\bibinfo {author} {\bibfnamefont {J.}~\bibnamefont
  {Kim}}, \bibinfo {author} {\bibfnamefont {J.}~\bibnamefont {Sinha}}, \bibinfo
  {author} {\bibfnamefont {M.}~\bibnamefont {Hayashi}}, \bibinfo {author}
  {\bibfnamefont {M.}~\bibnamefont {Yamanouchi}}, \bibinfo {author}
  {\bibfnamefont {S.}~\bibnamefont {Fukami}}, \bibinfo {author} {\bibfnamefont
  {T.}~\bibnamefont {Suzuki}}, \bibinfo {author} {\bibfnamefont
  {S.}~\bibnamefont {Mitani}},\ and\ \bibinfo {author} {\bibfnamefont
  {H.}~\bibnamefont {Ohno}},\ }\href {https://doi.org/10.1038/nmat3522}
  {\bibfield  {journal} {\bibinfo  {journal} {Nature Materials}\ }\textbf
  {\bibinfo {volume} {12}},\ \bibinfo {pages} {240} (\bibinfo {year}
  {2013})}\BibitemShut {NoStop}%
\bibitem [{\citenamefont {Garello}\ \emph {et~al.}(2013)\citenamefont
  {Garello}, \citenamefont {Miron}, \citenamefont {Avci}, \citenamefont
  {Freimuth}, \citenamefont {Mokrousov}, \citenamefont {Bl{\"u}gel},
  \citenamefont {Auffret}, \citenamefont {Boulle}, \citenamefont {Gaudin},\
  and\ \citenamefont {Gambardella}}]{Garello13}%
  \BibitemOpen
  \bibfield  {author} {\bibinfo {author} {\bibfnamefont {K.}~\bibnamefont
  {Garello}}, \bibinfo {author} {\bibfnamefont {I.~M.}\ \bibnamefont {Miron}},
  \bibinfo {author} {\bibfnamefont {C.~O.}\ \bibnamefont {Avci}}, \bibinfo
  {author} {\bibfnamefont {F.}~\bibnamefont {Freimuth}}, \bibinfo {author}
  {\bibfnamefont {Y.}~\bibnamefont {Mokrousov}}, \bibinfo {author}
  {\bibfnamefont {S.}~\bibnamefont {Bl{\"u}gel}}, \bibinfo {author}
  {\bibfnamefont {S.}~\bibnamefont {Auffret}}, \bibinfo {author} {\bibfnamefont
  {O.}~\bibnamefont {Boulle}}, \bibinfo {author} {\bibfnamefont
  {G.}~\bibnamefont {Gaudin}},\ and\ \bibinfo {author} {\bibfnamefont
  {P.}~\bibnamefont {Gambardella}},\ }\href
  {https://doi.org/10.1038/nnano.2013.145} {\bibfield  {journal} {\bibinfo
  {journal} {Nature Nanotechnology}\ }\textbf {\bibinfo {volume} {8}},\
  \bibinfo {pages} {587} (\bibinfo {year} {2013})}\BibitemShut {NoStop}%
\bibitem [{\citenamefont {Avci}\ \emph
  {et~al.}(2014{\natexlab{a}})\citenamefont {Avci}, \citenamefont {Garello},
  \citenamefont {Nistor}, \citenamefont {Godey}, \citenamefont {Ballesteros},
  \citenamefont {Mugarza}, \citenamefont {Barla}, \citenamefont {Valvidares},
  \citenamefont {Pellegrin}, \citenamefont {Ghosh}, \citenamefont {Miron},
  \citenamefont {Boulle}, \citenamefont {Auffret}, \citenamefont {Gaudin},\
  and\ \citenamefont {Gambardella}}]{Avci14}%
  \BibitemOpen
  \bibfield  {author} {\bibinfo {author} {\bibfnamefont {C.~O.}\ \bibnamefont
  {Avci}}, \bibinfo {author} {\bibfnamefont {K.}~\bibnamefont {Garello}},
  \bibinfo {author} {\bibfnamefont {C.}~\bibnamefont {Nistor}}, \bibinfo
  {author} {\bibfnamefont {S.}~\bibnamefont {Godey}}, \bibinfo {author}
  {\bibfnamefont {B.}~\bibnamefont {Ballesteros}}, \bibinfo {author}
  {\bibfnamefont {A.}~\bibnamefont {Mugarza}}, \bibinfo {author} {\bibfnamefont
  {A.}~\bibnamefont {Barla}}, \bibinfo {author} {\bibfnamefont
  {M.}~\bibnamefont {Valvidares}}, \bibinfo {author} {\bibfnamefont
  {E.}~\bibnamefont {Pellegrin}}, \bibinfo {author} {\bibfnamefont
  {A.}~\bibnamefont {Ghosh}}, \bibinfo {author} {\bibfnamefont {I.~M.}\
  \bibnamefont {Miron}}, \bibinfo {author} {\bibfnamefont {O.}~\bibnamefont
  {Boulle}}, \bibinfo {author} {\bibfnamefont {S.}~\bibnamefont {Auffret}},
  \bibinfo {author} {\bibfnamefont {G.}~\bibnamefont {Gaudin}},\ and\ \bibinfo
  {author} {\bibfnamefont {P.}~\bibnamefont {Gambardella}},\ }\href
  {https://doi.org/10.1103/PhysRevB.89.214419} {\bibfield  {journal} {\bibinfo
  {journal} {Phys. Rev. B}\ }\textbf {\bibinfo {volume} {89}},\ \bibinfo
  {pages} {214419} (\bibinfo {year} {2014}{\natexlab{a}})}\BibitemShut
  {NoStop}%
\bibitem [{\citenamefont {Avci}\ \emph
  {et~al.}(2014{\natexlab{b}})\citenamefont {Avci}, \citenamefont {Garello},
  \citenamefont {Gabureac}, \citenamefont {Ghosh}, \citenamefont {Fuhrer},
  \citenamefont {Alvarado},\ and\ \citenamefont {Gambardella}}]{Avci14b}%
  \BibitemOpen
  \bibfield  {author} {\bibinfo {author} {\bibfnamefont {C.~O.}\ \bibnamefont
  {Avci}}, \bibinfo {author} {\bibfnamefont {K.}~\bibnamefont {Garello}},
  \bibinfo {author} {\bibfnamefont {M.}~\bibnamefont {Gabureac}}, \bibinfo
  {author} {\bibfnamefont {A.}~\bibnamefont {Ghosh}}, \bibinfo {author}
  {\bibfnamefont {A.}~\bibnamefont {Fuhrer}}, \bibinfo {author} {\bibfnamefont
  {S.~F.}\ \bibnamefont {Alvarado}},\ and\ \bibinfo {author} {\bibfnamefont
  {P.}~\bibnamefont {Gambardella}},\ }\href
  {https://doi.org/10.1103/PhysRevB.90.224427} {\bibfield  {journal} {\bibinfo
  {journal} {Phys. Rev. B}\ }\textbf {\bibinfo {volume} {90}},\ \bibinfo
  {pages} {224427} (\bibinfo {year} {2014}{\natexlab{b}})}\BibitemShut
  {NoStop}%
\bibitem [{\citenamefont {Woo}\ \emph {et~al.}(2014)\citenamefont {Woo},
  \citenamefont {Mann}, \citenamefont {Tan}, \citenamefont {Caretta},\ and\
  \citenamefont {Beach}}]{Woo14}%
  \BibitemOpen
  \bibfield  {author} {\bibinfo {author} {\bibfnamefont {S.}~\bibnamefont
  {Woo}}, \bibinfo {author} {\bibfnamefont {M.}~\bibnamefont {Mann}}, \bibinfo
  {author} {\bibfnamefont {A.~J.}\ \bibnamefont {Tan}}, \bibinfo {author}
  {\bibfnamefont {L.}~\bibnamefont {Caretta}},\ and\ \bibinfo {author}
  {\bibfnamefont {G.~S.~D.}\ \bibnamefont {Beach}},\ }\href
  {https://doi.org/10.1063/1.4902529} {\bibfield  {journal} {\bibinfo
  {journal} {Applied Physics Letters}\ }\textbf {\bibinfo {volume} {105}},\
  \bibinfo {pages} {212404} (\bibinfo {year} {2014})}\BibitemShut {NoStop}%
\bibitem [{\citenamefont {Nguyen}\ \emph
  {et~al.}(2016{\natexlab{a}})\citenamefont {Nguyen}, \citenamefont {Ralph},\
  and\ \citenamefont {Buhrman}}]{Nguyen16}%
  \BibitemOpen
  \bibfield  {author} {\bibinfo {author} {\bibfnamefont {M.-H.}\ \bibnamefont
  {Nguyen}}, \bibinfo {author} {\bibfnamefont {D.~C.}\ \bibnamefont {Ralph}},\
  and\ \bibinfo {author} {\bibfnamefont {R.~A.}\ \bibnamefont {Buhrman}},\
  }\href {https://doi.org/10.1103/PhysRevLett.116.126601} {\bibfield  {journal}
  {\bibinfo  {journal} {Phys. Rev. Lett.}\ }\textbf {\bibinfo {volume} {116}},\
  \bibinfo {pages} {126601} (\bibinfo {year} {2016}{\natexlab{a}})}\BibitemShut
  {NoStop}%
\bibitem [{\citenamefont {Nguyen}\ \emph
  {et~al.}(2016{\natexlab{b}})\citenamefont {Nguyen}, \citenamefont {Zhao},
  \citenamefont {Ralph},\ and\ \citenamefont {Buhrman}}]{Nguyen16b}%
  \BibitemOpen
  \bibfield  {author} {\bibinfo {author} {\bibfnamefont {M.-H.}\ \bibnamefont
  {Nguyen}}, \bibinfo {author} {\bibfnamefont {M.}~\bibnamefont {Zhao}},
  \bibinfo {author} {\bibfnamefont {D.~C.}\ \bibnamefont {Ralph}},\ and\
  \bibinfo {author} {\bibfnamefont {R.~A.}\ \bibnamefont {Buhrman}},\ }\href
  {https://doi.org/10.1063/1.4953768} {\bibfield  {journal} {\bibinfo
  {journal} {Applied Physics Letters}\ }\textbf {\bibinfo {volume} {108}},\
  \bibinfo {pages} {242407} (\bibinfo {year} {2016}{\natexlab{b}})}\BibitemShut
  {NoStop}%
\bibitem [{\citenamefont {Akyol}\ \emph {et~al.}(2016)\citenamefont {Akyol},
  \citenamefont {Jiang}, \citenamefont {Yu}, \citenamefont {Fan}, \citenamefont
  {Gunes}, \citenamefont {Ekicibil}, \citenamefont {Khalili~Amiri},\ and\
  \citenamefont {Wang}}]{Aykol16}%
  \BibitemOpen
  \bibfield  {author} {\bibinfo {author} {\bibfnamefont {M.}~\bibnamefont
  {Akyol}}, \bibinfo {author} {\bibfnamefont {W.}~\bibnamefont {Jiang}},
  \bibinfo {author} {\bibfnamefont {G.}~\bibnamefont {Yu}}, \bibinfo {author}
  {\bibfnamefont {Y.}~\bibnamefont {Fan}}, \bibinfo {author} {\bibfnamefont
  {M.}~\bibnamefont {Gunes}}, \bibinfo {author} {\bibfnamefont
  {A.}~\bibnamefont {Ekicibil}}, \bibinfo {author} {\bibfnamefont
  {P.}~\bibnamefont {Khalili~Amiri}},\ and\ \bibinfo {author} {\bibfnamefont
  {K.~L.}\ \bibnamefont {Wang}},\ }\href {https://doi.org/10.1063/1.4958295}
  {\bibfield  {journal} {\bibinfo  {journal} {Applied Physics Letters}\
  }\textbf {\bibinfo {volume} {109}},\ \bibinfo {pages} {022403} (\bibinfo
  {year} {2016})}\BibitemShut {NoStop}%
\bibitem [{\citenamefont {Ou}\ \emph {et~al.}(2016)\citenamefont {Ou},
  \citenamefont {Pai}, \citenamefont {Shi}, \citenamefont {Ralph},\ and\
  \citenamefont {Buhrman}}]{Ou16b}%
  \BibitemOpen
  \bibfield  {author} {\bibinfo {author} {\bibfnamefont {Y.}~\bibnamefont
  {Ou}}, \bibinfo {author} {\bibfnamefont {C.-F.}\ \bibnamefont {Pai}},
  \bibinfo {author} {\bibfnamefont {S.}~\bibnamefont {Shi}}, \bibinfo {author}
  {\bibfnamefont {D.~C.}\ \bibnamefont {Ralph}},\ and\ \bibinfo {author}
  {\bibfnamefont {R.~A.}\ \bibnamefont {Buhrman}},\ }\href
  {https://doi.org/10.1103/PhysRevB.94.140414} {\bibfield  {journal} {\bibinfo
  {journal} {Phys. Rev. B}\ }\textbf {\bibinfo {volume} {94}},\ \bibinfo
  {pages} {140414} (\bibinfo {year} {2016})}\BibitemShut {NoStop}%
\bibitem [{\citenamefont {Hayashi}\ \emph {et~al.}(2014)\citenamefont
  {Hayashi}, \citenamefont {Kim}, \citenamefont {Yamanouchi},\ and\
  \citenamefont {Ohno}}]{Hayashi14}%
  \BibitemOpen
  \bibfield  {author} {\bibinfo {author} {\bibfnamefont {M.}~\bibnamefont
  {Hayashi}}, \bibinfo {author} {\bibfnamefont {J.}~\bibnamefont {Kim}},
  \bibinfo {author} {\bibfnamefont {M.}~\bibnamefont {Yamanouchi}},\ and\
  \bibinfo {author} {\bibfnamefont {H.}~\bibnamefont {Ohno}},\ }\href
  {https://doi.org/10.1103/PhysRevB.89.144425} {\bibfield  {journal} {\bibinfo
  {journal} {Phys. Rev. B}\ }\textbf {\bibinfo {volume} {89}},\ \bibinfo
  {pages} {144425} (\bibinfo {year} {2014})}\BibitemShut {NoStop}%
\bibitem [{\citenamefont {Kondou}\ \emph {et~al.}(2016)\citenamefont {Kondou},
  \citenamefont {Sukegawa}, \citenamefont {Kasai}, \citenamefont {Mitani},
  \citenamefont {Niimi},\ and\ \citenamefont {Otani}}]{Kondou16}%
  \BibitemOpen
  \bibfield  {author} {\bibinfo {author} {\bibfnamefont {K.}~\bibnamefont
  {Kondou}}, \bibinfo {author} {\bibfnamefont {H.}~\bibnamefont {Sukegawa}},
  \bibinfo {author} {\bibfnamefont {S.}~\bibnamefont {Kasai}}, \bibinfo
  {author} {\bibfnamefont {S.}~\bibnamefont {Mitani}}, \bibinfo {author}
  {\bibfnamefont {Y.}~\bibnamefont {Niimi}},\ and\ \bibinfo {author}
  {\bibfnamefont {Y.}~\bibnamefont {Otani}},\ }\href
  {https://doi.org/10.7567/apex.9.023002} {\bibfield  {journal} {\bibinfo
  {journal} {Applied Physics Express}\ }\textbf {\bibinfo {volume} {9}},\
  \bibinfo {pages} {023002} (\bibinfo {year} {2016})}\BibitemShut {NoStop}%
\bibitem [{\citenamefont {Vlietstra}\ \emph {et~al.}(2014)\citenamefont
  {Vlietstra}, \citenamefont {Shan}, \citenamefont {van Wees}, \citenamefont
  {Isasa}, \citenamefont {Casanova},\ and\ \citenamefont
  {Ben~Youssef}}]{Vliestra14}%
  \BibitemOpen
  \bibfield  {author} {\bibinfo {author} {\bibfnamefont {N.}~\bibnamefont
  {Vlietstra}}, \bibinfo {author} {\bibfnamefont {J.}~\bibnamefont {Shan}},
  \bibinfo {author} {\bibfnamefont {B.~J.}\ \bibnamefont {van Wees}}, \bibinfo
  {author} {\bibfnamefont {M.}~\bibnamefont {Isasa}}, \bibinfo {author}
  {\bibfnamefont {F.}~\bibnamefont {Casanova}},\ and\ \bibinfo {author}
  {\bibfnamefont {J.}~\bibnamefont {Ben~Youssef}},\ }\href
  {https://doi.org/10.1103/PhysRevB.90.174436} {\bibfield  {journal} {\bibinfo
  {journal} {Phys. Rev. B}\ }\textbf {\bibinfo {volume} {90}},\ \bibinfo
  {pages} {174436} (\bibinfo {year} {2014})}\BibitemShut {NoStop}%
\bibitem [{\citenamefont {Liu}\ \emph {et~al.}(2021)\citenamefont {Liu},
  \citenamefont {Zhang}, \citenamefont {Sun}, \citenamefont {Miao},
  \citenamefont {Wang},\ and\ \citenamefont {Ding}}]{Liu21}%
  \BibitemOpen
  \bibfield  {author} {\bibinfo {author} {\bibfnamefont {Q.}~\bibnamefont
  {Liu}}, \bibinfo {author} {\bibfnamefont {Y.}~\bibnamefont {Zhang}}, \bibinfo
  {author} {\bibfnamefont {L.}~\bibnamefont {Sun}}, \bibinfo {author}
  {\bibfnamefont {B.}~\bibnamefont {Miao}}, \bibinfo {author} {\bibfnamefont
  {X.~R.}\ \bibnamefont {Wang}},\ and\ \bibinfo {author} {\bibfnamefont
  {H.~F.}\ \bibnamefont {Ding}},\ }\href {https://doi.org/10.1063/5.0038567}
  {\bibfield  {journal} {\bibinfo  {journal} {Applied Physics Letters}\
  }\textbf {\bibinfo {volume} {118}},\ \bibinfo {pages} {132401} (\bibinfo
  {year} {2021})},\ \Eprint
  {https://arxiv.org/abs/https://doi.org/10.1063/5.0038567}
  {https://doi.org/10.1063/5.0038567} \BibitemShut {NoStop}%
\bibitem [{\citenamefont {Montazeri}\ \emph {et~al.}(2015)\citenamefont
  {Montazeri}, \citenamefont {Upadhyaya}, \citenamefont {Onbasli},
  \citenamefont {Yu}, \citenamefont {Wong}, \citenamefont {Lang}, \citenamefont
  {Fan}, \citenamefont {Li}, \citenamefont {Khalili~Amiri}, \citenamefont
  {Schwartz}, \citenamefont {Ross},\ and\ \citenamefont {Wang}}]{Montazeri15}%
  \BibitemOpen
  \bibfield  {author} {\bibinfo {author} {\bibfnamefont {M.}~\bibnamefont
  {Montazeri}}, \bibinfo {author} {\bibfnamefont {P.}~\bibnamefont
  {Upadhyaya}}, \bibinfo {author} {\bibfnamefont {M.~C.}\ \bibnamefont
  {Onbasli}}, \bibinfo {author} {\bibfnamefont {G.}~\bibnamefont {Yu}},
  \bibinfo {author} {\bibfnamefont {K.~L.}\ \bibnamefont {Wong}}, \bibinfo
  {author} {\bibfnamefont {M.}~\bibnamefont {Lang}}, \bibinfo {author}
  {\bibfnamefont {Y.}~\bibnamefont {Fan}}, \bibinfo {author} {\bibfnamefont
  {X.}~\bibnamefont {Li}}, \bibinfo {author} {\bibfnamefont {P.}~\bibnamefont
  {Khalili~Amiri}}, \bibinfo {author} {\bibfnamefont {R.~N.}\ \bibnamefont
  {Schwartz}}, \bibinfo {author} {\bibfnamefont {C.~A.}\ \bibnamefont {Ross}},\
  and\ \bibinfo {author} {\bibfnamefont {K.~L.}\ \bibnamefont {Wang}},\ }\href
  {https://doi.org/10.1038/ncomms9958} {\bibfield  {journal} {\bibinfo
  {journal} {Nature Communications}\ }\textbf {\bibinfo {volume} {6}},\
  \bibinfo {pages} {8958} (\bibinfo {year} {2015})}\BibitemShut {NoStop}%
\bibitem [{\citenamefont {McGuire}\ and\ \citenamefont
  {Potter}(1975)}]{McGuire75}%
  \BibitemOpen
  \bibfield  {author} {\bibinfo {author} {\bibfnamefont {T.}~\bibnamefont
  {McGuire}}\ and\ \bibinfo {author} {\bibfnamefont {R.}~\bibnamefont
  {Potter}},\ }\href {https://doi.org/10.1109/TMAG.1975.1058782} {\bibfield
  {journal} {\bibinfo  {journal} {IEEE Transactions on Magnetics}\ }\textbf
  {\bibinfo {volume} {11}},\ \bibinfo {pages} {1018} (\bibinfo {year}
  {1975})}\BibitemShut {NoStop}%
\bibitem [{\citenamefont {Fan}\ \emph {et~al.}(2016)\citenamefont {Fan},
  \citenamefont {Mellnik}, \citenamefont {Wang}, \citenamefont {Reynolds},
  \citenamefont {Wang}, \citenamefont {Celik}, \citenamefont {Lorenz},
  \citenamefont {Ralph},\ and\ \citenamefont {Xiao}}]{Fan16}%
  \BibitemOpen
  \bibfield  {author} {\bibinfo {author} {\bibfnamefont {X.}~\bibnamefont
  {Fan}}, \bibinfo {author} {\bibfnamefont {A.~R.}\ \bibnamefont {Mellnik}},
  \bibinfo {author} {\bibfnamefont {W.}~\bibnamefont {Wang}}, \bibinfo {author}
  {\bibfnamefont {N.}~\bibnamefont {Reynolds}}, \bibinfo {author}
  {\bibfnamefont {T.}~\bibnamefont {Wang}}, \bibinfo {author} {\bibfnamefont
  {H.}~\bibnamefont {Celik}}, \bibinfo {author} {\bibfnamefont {V.~O.}\
  \bibnamefont {Lorenz}}, \bibinfo {author} {\bibfnamefont {D.~C.}\
  \bibnamefont {Ralph}},\ and\ \bibinfo {author} {\bibfnamefont {J.~Q.}\
  \bibnamefont {Xiao}},\ }\href {https://doi.org/10.1063/1.4962402} {\bibfield
  {journal} {\bibinfo  {journal} {Applied Physics Letters}\ }\textbf {\bibinfo
  {volume} {109}},\ \bibinfo {pages} {122406} (\bibinfo {year}
  {2016})}\BibitemShut {NoStop}%
\bibitem [{\citenamefont {Kim}\ \emph {et~al.}(2019)\citenamefont {Kim},
  \citenamefont {Park}, \citenamefont {Whang}, \citenamefont {Park},
  \citenamefont {Min},\ and\ \citenamefont {Choe}}]{Kim19}%
  \BibitemOpen
  \bibfield  {author} {\bibinfo {author} {\bibfnamefont {J.-S.}\ \bibnamefont
  {Kim}}, \bibinfo {author} {\bibfnamefont {Y.-K.}\ \bibnamefont {Park}},
  \bibinfo {author} {\bibfnamefont {H.-S.}\ \bibnamefont {Whang}}, \bibinfo
  {author} {\bibfnamefont {J.-H.}\ \bibnamefont {Park}}, \bibinfo {author}
  {\bibfnamefont {B.-C.}\ \bibnamefont {Min}},\ and\ \bibinfo {author}
  {\bibfnamefont {S.-B.}\ \bibnamefont {Choe}},\ }\href
  {https://doi.org/10.1063/1.5087743} {\bibfield  {journal} {\bibinfo
  {journal} {Applied Physics Letters}\ }\textbf {\bibinfo {volume} {114}},\
  \bibinfo {pages} {182402} (\bibinfo {year} {2019})},\ \Eprint
  {https://arxiv.org/abs/https://doi.org/10.1063/1.5087743}
  {https://doi.org/10.1063/1.5087743} \BibitemShut {NoStop}%
\bibitem [{\citenamefont {{Celik}}\ \emph {et~al.}(2019)\citenamefont
  {{Celik}}, \citenamefont {{Kannan}}, \citenamefont {{Wang}}, \citenamefont
  {{Mellnik}}, \citenamefont {{Fan}}, \citenamefont {{Zhou}}, \citenamefont
  {{Barri}}, \citenamefont {{Ralph}}, \citenamefont {{Doty}}, \citenamefont
  {{Lorenz}},\ and\ \citenamefont {{Xiao}}}]{Celik19}%
  \BibitemOpen
  \bibfield  {author} {\bibinfo {author} {\bibfnamefont {H.}~\bibnamefont
  {{Celik}}}, \bibinfo {author} {\bibfnamefont {H.}~\bibnamefont {{Kannan}}},
  \bibinfo {author} {\bibfnamefont {T.}~\bibnamefont {{Wang}}}, \bibinfo
  {author} {\bibfnamefont {A.~R.}\ \bibnamefont {{Mellnik}}}, \bibinfo {author}
  {\bibfnamefont {X.}~\bibnamefont {{Fan}}}, \bibinfo {author} {\bibfnamefont
  {X.}~\bibnamefont {{Zhou}}}, \bibinfo {author} {\bibfnamefont
  {R.}~\bibnamefont {{Barri}}}, \bibinfo {author} {\bibfnamefont {D.~C.}\
  \bibnamefont {{Ralph}}}, \bibinfo {author} {\bibfnamefont {M.~F.}\
  \bibnamefont {{Doty}}}, \bibinfo {author} {\bibfnamefont {V.~O.}\
  \bibnamefont {{Lorenz}}},\ and\ \bibinfo {author} {\bibfnamefont {J.~Q.}\
  \bibnamefont {{Xiao}}},\ }\href {https://doi.org/10.1109/TMAG.2018.2873129}
  {\bibfield  {journal} {\bibinfo  {journal} {IEEE Transactions on Magnetics}\
  }\textbf {\bibinfo {volume} {55}},\ \bibinfo {pages} {1} (\bibinfo {year}
  {2019})}\BibitemShut {NoStop}%
\bibitem [{\citenamefont {Xing}\ \emph {et~al.}(2020)\citenamefont {Xing},
  \citenamefont {Zhou}, \citenamefont {Wang}, \citenamefont {Li}, \citenamefont
  {Cao}, \citenamefont {Cai}, \citenamefont {Zhang}, \citenamefont {Ji},
  \citenamefont {Lin}, \citenamefont {Wu}, \citenamefont {Lei}, \citenamefont
  {Zhang},\ and\ \citenamefont {Zhao}}]{Xing20}%
  \BibitemOpen
  \bibfield  {author} {\bibinfo {author} {\bibfnamefont {T.}~\bibnamefont
  {Xing}}, \bibinfo {author} {\bibfnamefont {C.}~\bibnamefont {Zhou}}, \bibinfo
  {author} {\bibfnamefont {C.~X.}\ \bibnamefont {Wang}}, \bibinfo {author}
  {\bibfnamefont {Z.}~\bibnamefont {Li}}, \bibinfo {author} {\bibfnamefont
  {A.~N.}\ \bibnamefont {Cao}}, \bibinfo {author} {\bibfnamefont {W.~L.}\
  \bibnamefont {Cai}}, \bibinfo {author} {\bibfnamefont {X.~Y.}\ \bibnamefont
  {Zhang}}, \bibinfo {author} {\bibfnamefont {B.}~\bibnamefont {Ji}}, \bibinfo
  {author} {\bibfnamefont {T.}~\bibnamefont {Lin}}, \bibinfo {author}
  {\bibfnamefont {Y.~Z.}\ \bibnamefont {Wu}}, \bibinfo {author} {\bibfnamefont
  {N.}~\bibnamefont {Lei}}, \bibinfo {author} {\bibfnamefont {Y.~G.}\
  \bibnamefont {Zhang}},\ and\ \bibinfo {author} {\bibfnamefont {W.~S.}\
  \bibnamefont {Zhao}},\ }\href {https://doi.org/10.1103/PhysRevB.101.224407}
  {\bibfield  {journal} {\bibinfo  {journal} {Phys. Rev. B}\ }\textbf {\bibinfo
  {volume} {101}},\ \bibinfo {pages} {224407} (\bibinfo {year}
  {2020})}\BibitemShut {NoStop}%
\bibitem [{\citenamefont {Nguyen}\ and\ \citenamefont {Pai}(2021)}]{Nguyen21}%
  \BibitemOpen
  \bibfield  {author} {\bibinfo {author} {\bibfnamefont {M.-H.}\ \bibnamefont
  {Nguyen}}\ and\ \bibinfo {author} {\bibfnamefont {C.-F.}\ \bibnamefont
  {Pai}},\ }\href {https://doi.org/10.1063/5.0041123} {\bibfield  {journal}
  {\bibinfo  {journal} {APL Materials}\ }\textbf {\bibinfo {volume} {9}},\
  \bibinfo {pages} {030902} (\bibinfo {year} {2021})},\ \Eprint
  {https://arxiv.org/abs/https://doi.org/10.1063/5.0041123}
  {https://doi.org/10.1063/5.0041123} \BibitemShut {NoStop}%
\bibitem [{\citenamefont {Postava}\ \emph {et~al.}(2002)\citenamefont
  {Postava}, \citenamefont {Hrabovsky}, \citenamefont {Pistora}, \citenamefont
  {Fert}, \citenamefont {Visnovsky},\ and\ \citenamefont
  {Yamaguchi}}]{Postava02}%
  \BibitemOpen
  \bibfield  {author} {\bibinfo {author} {\bibfnamefont {K.}~\bibnamefont
  {Postava}}, \bibinfo {author} {\bibfnamefont {D.}~\bibnamefont {Hrabovsky}},
  \bibinfo {author} {\bibfnamefont {J.}~\bibnamefont {Pistora}}, \bibinfo
  {author} {\bibfnamefont {A.~R.}\ \bibnamefont {Fert}}, \bibinfo {author}
  {\bibfnamefont {S.}~\bibnamefont {Visnovsky}},\ and\ \bibinfo {author}
  {\bibfnamefont {T.}~\bibnamefont {Yamaguchi}},\ }\href
  {https://doi.org/10.1063/1.1449436} {\bibfield  {journal} {\bibinfo
  {journal} {Journal of Applied Physics}\ }\textbf {\bibinfo {volume} {91}},\
  \bibinfo {pages} {7293} (\bibinfo {year} {2002})},\ \Eprint
  {https://arxiv.org/abs/https://aip.scitation.org/doi/pdf/10.1063/1.1449436}
  {https://aip.scitation.org/doi/pdf/10.1063/1.1449436} \BibitemShut {NoStop}%
\bibitem [{\citenamefont {Silber}\ \emph {et~al.}(2019)\citenamefont {Silber},
  \citenamefont {Stejskal}, \citenamefont {Beran}, \citenamefont {Cejpek},
  \citenamefont {Antos}, \citenamefont {Matalla-Wagner}, \citenamefont {Thien},
  \citenamefont {Kuschel}, \citenamefont {Wollschlager}, \citenamefont {Veis},
  \citenamefont {Kuschel},\ and\ \citenamefont {Hamrle}}]{Silver19}%
  \BibitemOpen
  \bibfield  {author} {\bibinfo {author} {\bibfnamefont {R.}~\bibnamefont
  {Silber}}, \bibinfo {author} {\bibfnamefont {O.}~\bibnamefont {Stejskal}},
  \bibinfo {author} {\bibfnamefont {L.}~\bibnamefont {Beran}}, \bibinfo
  {author} {\bibfnamefont {P.}~\bibnamefont {Cejpek}}, \bibinfo {author}
  {\bibfnamefont {R.}~\bibnamefont {Antos}}, \bibinfo {author} {\bibfnamefont
  {T.}~\bibnamefont {Matalla-Wagner}}, \bibinfo {author} {\bibfnamefont
  {J.}~\bibnamefont {Thien}}, \bibinfo {author} {\bibfnamefont
  {O.}~\bibnamefont {Kuschel}}, \bibinfo {author} {\bibfnamefont
  {J.}~\bibnamefont {Wollschlager}}, \bibinfo {author} {\bibfnamefont
  {M.}~\bibnamefont {Veis}}, \bibinfo {author} {\bibfnamefont {T.}~\bibnamefont
  {Kuschel}},\ and\ \bibinfo {author} {\bibfnamefont {J.}~\bibnamefont
  {Hamrle}},\ }\href {https://doi.org/10.1103/PhysRevB.100.064403} {\bibfield
  {journal} {\bibinfo  {journal} {Phys. Rev. B}\ }\textbf {\bibinfo {volume}
  {100}},\ \bibinfo {pages} {064403} (\bibinfo {year} {2019})}\BibitemShut
  {NoStop}%
\bibitem [{\citenamefont {Qiu}\ and\ \citenamefont {Bader}(2000)}]{Qiu00}%
  \BibitemOpen
  \bibfield  {author} {\bibinfo {author} {\bibfnamefont {Z.~Q.}\ \bibnamefont
  {Qiu}}\ and\ \bibinfo {author} {\bibfnamefont {S.~D.}\ \bibnamefont
  {Bader}},\ }\href {https://doi.org/10.1063/1.1150496} {\bibfield  {journal}
  {\bibinfo  {journal} {Review of Scientific Instruments}\ }\textbf {\bibinfo
  {volume} {71}},\ \bibinfo {pages} {1243} (\bibinfo {year} {2000})},\ \Eprint
  {https://arxiv.org/abs/https://doi.org/10.1063/1.1150496}
  {https://doi.org/10.1063/1.1150496} \BibitemShut {NoStop}%
\bibitem [{\citenamefont {Berger}\ and\ \citenamefont
  {Pufall}(1997)}]{Berger97}%
  \BibitemOpen
  \bibfield  {author} {\bibinfo {author} {\bibfnamefont {A.}~\bibnamefont
  {Berger}}\ and\ \bibinfo {author} {\bibfnamefont {M.~R.}\ \bibnamefont
  {Pufall}},\ }\href {https://doi.org/10.1063/1.119669} {\bibfield  {journal}
  {\bibinfo  {journal} {Applied Physics Letters}\ }\textbf {\bibinfo {volume}
  {71}},\ \bibinfo {pages} {965} (\bibinfo {year} {1997})},\ \Eprint
  {https://arxiv.org/abs/https://doi.org/10.1063/1.119669}
  {https://doi.org/10.1063/1.119669} \BibitemShut {NoStop}%
\bibitem [{\citenamefont {Gonzalez-Fuentes}\ \emph
  {et~al.}(2020{\natexlab{a}})\citenamefont {Gonzalez-Fuentes}, \citenamefont
  {Orellana}, \citenamefont {Romanque-Albornoz},\ and\ \citenamefont
  {Garcia}}]{Gonzalez-Fuentes19}%
  \BibitemOpen
  \bibfield  {author} {\bibinfo {author} {\bibfnamefont {C.}~\bibnamefont
  {Gonzalez-Fuentes}}, \bibinfo {author} {\bibfnamefont {C.}~\bibnamefont
  {Orellana}}, \bibinfo {author} {\bibfnamefont {C.}~\bibnamefont
  {Romanque-Albornoz}},\ and\ \bibinfo {author} {\bibfnamefont
  {C.}~\bibnamefont {Garcia}},\ }\href
  {https://doi.org/10.1109/TIM.2020.2987453} {\bibfield  {journal} {\bibinfo
  {journal} {IEEE Transactions on Instrumentation and Measurement}\ }\textbf
  {\bibinfo {volume} {69}},\ \bibinfo {pages} {8432} (\bibinfo {year}
  {2020}{\natexlab{a}})}\BibitemShut {NoStop}%
\bibitem [{\citenamefont {Higo}\ \emph {et~al.}(2018)\citenamefont {Higo},
  \citenamefont {Man}, \citenamefont {Gopman}, \citenamefont {Wu},
  \citenamefont {Koretsune}, \citenamefont {van~'t Erve}, \citenamefont
  {Kabanov}, \citenamefont {Rees}, \citenamefont {Li}, \citenamefont {Suzuki},
  \citenamefont {Patankar}, \citenamefont {Ikhlas}, \citenamefont {Chien},
  \citenamefont {Arita}, \citenamefont {Shull}, \citenamefont {Orenstein},\
  and\ \citenamefont {Nakatsuji}}]{Higo18}%
  \BibitemOpen
  \bibfield  {author} {\bibinfo {author} {\bibfnamefont {T.}~\bibnamefont
  {Higo}}, \bibinfo {author} {\bibfnamefont {H.}~\bibnamefont {Man}}, \bibinfo
  {author} {\bibfnamefont {D.~B.}\ \bibnamefont {Gopman}}, \bibinfo {author}
  {\bibfnamefont {L.}~\bibnamefont {Wu}}, \bibinfo {author} {\bibfnamefont
  {T.}~\bibnamefont {Koretsune}}, \bibinfo {author} {\bibfnamefont {O.~M.~J.}\
  \bibnamefont {van~'t Erve}}, \bibinfo {author} {\bibfnamefont {Y.~P.}\
  \bibnamefont {Kabanov}}, \bibinfo {author} {\bibfnamefont {D.}~\bibnamefont
  {Rees}}, \bibinfo {author} {\bibfnamefont {Y.}~\bibnamefont {Li}}, \bibinfo
  {author} {\bibfnamefont {M.-T.}\ \bibnamefont {Suzuki}}, \bibinfo {author}
  {\bibfnamefont {S.}~\bibnamefont {Patankar}}, \bibinfo {author}
  {\bibfnamefont {M.}~\bibnamefont {Ikhlas}}, \bibinfo {author} {\bibfnamefont
  {C.~L.}\ \bibnamefont {Chien}}, \bibinfo {author} {\bibfnamefont
  {R.}~\bibnamefont {Arita}}, \bibinfo {author} {\bibfnamefont {R.~D.}\
  \bibnamefont {Shull}}, \bibinfo {author} {\bibfnamefont {J.}~\bibnamefont
  {Orenstein}},\ and\ \bibinfo {author} {\bibfnamefont {S.}~\bibnamefont
  {Nakatsuji}},\ }\href {https://doi.org/10.1038/s41566-017-0086-z} {\bibfield
  {journal} {\bibinfo  {journal} {Nature Photonics}\ }\textbf {\bibinfo
  {volume} {12}},\ \bibinfo {pages} {73} (\bibinfo {year} {2018})}\BibitemShut
  {NoStop}%
\bibitem [{\citenamefont {Soya}\ \emph {et~al.}(2021)\citenamefont {Soya},
  \citenamefont {Hayashi}, \citenamefont {Harumoto}, \citenamefont {Gao},
  \citenamefont {Haku},\ and\ \citenamefont {Ando}}]{Soya21}%
  \BibitemOpen
  \bibfield  {author} {\bibinfo {author} {\bibfnamefont {N.}~\bibnamefont
  {Soya}}, \bibinfo {author} {\bibfnamefont {H.}~\bibnamefont {Hayashi}},
  \bibinfo {author} {\bibfnamefont {T.}~\bibnamefont {Harumoto}}, \bibinfo
  {author} {\bibfnamefont {T.}~\bibnamefont {Gao}}, \bibinfo {author}
  {\bibfnamefont {S.}~\bibnamefont {Haku}},\ and\ \bibinfo {author}
  {\bibfnamefont {K.}~\bibnamefont {Ando}},\ }\href
  {https://doi.org/10.1103/PhysRevB.103.174427} {\bibfield  {journal} {\bibinfo
   {journal} {Phys. Rev. B}\ }\textbf {\bibinfo {volume} {103}},\ \bibinfo
  {pages} {174427} (\bibinfo {year} {2021})}\BibitemShut {NoStop}%
\bibitem [{\citenamefont {Dutta}\ \emph {et~al.}(2021)\citenamefont {Dutta},
  \citenamefont {Bose}, \citenamefont {Tulapurkar}, \citenamefont {Buhrman},\
  and\ \citenamefont {Ralph}}]{Dutta21}%
  \BibitemOpen
  \bibfield  {author} {\bibinfo {author} {\bibfnamefont {S.}~\bibnamefont
  {Dutta}}, \bibinfo {author} {\bibfnamefont {A.}~\bibnamefont {Bose}},
  \bibinfo {author} {\bibfnamefont {A.~A.}\ \bibnamefont {Tulapurkar}},
  \bibinfo {author} {\bibfnamefont {R.~A.}\ \bibnamefont {Buhrman}},\ and\
  \bibinfo {author} {\bibfnamefont {D.~C.}\ \bibnamefont {Ralph}},\ }\href
  {https://doi.org/10.1103/PhysRevB.103.184416} {\bibfield  {journal} {\bibinfo
   {journal} {Phys. Rev. B}\ }\textbf {\bibinfo {volume} {103}},\ \bibinfo
  {pages} {184416} (\bibinfo {year} {2021})}\BibitemShut {NoStop}%
\bibitem [{\citenamefont {Gonzalez-Fuentes}\ \emph
  {et~al.}(2020{\natexlab{b}})\citenamefont {Gonzalez-Fuentes}, \citenamefont
  {Orellana}, \citenamefont {Romanque-Albornoz},\ and\ \citenamefont
  {Garcia}}]{Gonzalez-Fuentes20}%
  \BibitemOpen
  \bibfield  {author} {\bibinfo {author} {\bibfnamefont {C.}~\bibnamefont
  {Gonzalez-Fuentes}}, \bibinfo {author} {\bibfnamefont {C.}~\bibnamefont
  {Orellana}}, \bibinfo {author} {\bibfnamefont {C.}~\bibnamefont
  {Romanque-Albornoz}},\ and\ \bibinfo {author} {\bibfnamefont
  {C.}~\bibnamefont {Garcia}},\ }\href
  {https://doi.org/10.1109/TIM.2020.2987453} {\bibfield  {journal} {\bibinfo
  {journal} {IEEE Transactions on Instrumentation and Measurement}\ }\textbf
  {\bibinfo {volume} {69}},\ \bibinfo {pages} {8432} (\bibinfo {year}
  {2020}{\natexlab{b}})}\BibitemShut {NoStop}%
\bibitem [{\citenamefont {Boone}\ \emph {et~al.}(2015)\citenamefont {Boone},
  \citenamefont {Shaw}, \citenamefont {Nembach},\ and\ \citenamefont
  {Silva}}]{Boone15}%
  \BibitemOpen
  \bibfield  {author} {\bibinfo {author} {\bibfnamefont {C.~T.}\ \bibnamefont
  {Boone}}, \bibinfo {author} {\bibfnamefont {J.~M.}\ \bibnamefont {Shaw}},
  \bibinfo {author} {\bibfnamefont {H.~T.}\ \bibnamefont {Nembach}},\ and\
  \bibinfo {author} {\bibfnamefont {T.~J.}\ \bibnamefont {Silva}},\ }\href
  {https://doi.org/10.1063/1.4922581} {\bibfield  {journal} {\bibinfo
  {journal} {Journal of Applied Physics}\ }\textbf {\bibinfo {volume} {117}},\
  \bibinfo {pages} {223910} (\bibinfo {year} {2015})}\BibitemShut {NoStop}%
\bibitem [{\citenamefont {Fan}\ \emph {et~al.}(2013)\citenamefont {Fan},
  \citenamefont {Wu}, \citenamefont {Chen}, \citenamefont {Jerry},
  \citenamefont {Zhang},\ and\ \citenamefont {Xiao}}]{Fan13}%
  \BibitemOpen
  \bibfield  {author} {\bibinfo {author} {\bibfnamefont {X.}~\bibnamefont
  {Fan}}, \bibinfo {author} {\bibfnamefont {J.}~\bibnamefont {Wu}}, \bibinfo
  {author} {\bibfnamefont {Y.}~\bibnamefont {Chen}}, \bibinfo {author}
  {\bibfnamefont {M.~J.}\ \bibnamefont {Jerry}}, \bibinfo {author}
  {\bibfnamefont {H.}~\bibnamefont {Zhang}},\ and\ \bibinfo {author}
  {\bibfnamefont {J.~Q.}\ \bibnamefont {Xiao}},\ }\href
  {https://doi.org/10.1038/ncomms2709} {\bibfield  {journal} {\bibinfo
  {journal} {Nature Communications}\ }\textbf {\bibinfo {volume} {4}},\
  \bibinfo {pages} {1799} (\bibinfo {year} {2013})}\BibitemShut {NoStop}%
\bibitem [{\citenamefont {Emori}\ \emph {et~al.}(2016)\citenamefont {Emori},
  \citenamefont {Nan}, \citenamefont {Belkessam}, \citenamefont {Wang},
  \citenamefont {Matyushov}, \citenamefont {Babroski}, \citenamefont {Gao},
  \citenamefont {Lin},\ and\ \citenamefont {Sun}}]{Emori16}%
  \BibitemOpen
  \bibfield  {author} {\bibinfo {author} {\bibfnamefont {S.}~\bibnamefont
  {Emori}}, \bibinfo {author} {\bibfnamefont {T.}~\bibnamefont {Nan}}, \bibinfo
  {author} {\bibfnamefont {A.~M.}\ \bibnamefont {Belkessam}}, \bibinfo {author}
  {\bibfnamefont {X.}~\bibnamefont {Wang}}, \bibinfo {author} {\bibfnamefont
  {A.~D.}\ \bibnamefont {Matyushov}}, \bibinfo {author} {\bibfnamefont {C.~J.}\
  \bibnamefont {Babroski}}, \bibinfo {author} {\bibfnamefont {Y.}~\bibnamefont
  {Gao}}, \bibinfo {author} {\bibfnamefont {H.}~\bibnamefont {Lin}},\ and\
  \bibinfo {author} {\bibfnamefont {N.~X.}\ \bibnamefont {Sun}},\ }\href
  {https://doi.org/10.1103/PhysRevB.93.180402} {\bibfield  {journal} {\bibinfo
  {journal} {Phys. Rev. B}\ }\textbf {\bibinfo {volume} {93}},\ \bibinfo
  {pages} {180402} (\bibinfo {year} {2016})}\BibitemShut {NoStop}%
\bibitem [{\citenamefont {Gonzalez-Fuentes}\ \emph {et~al.}(2021)\citenamefont
  {Gonzalez-Fuentes}, \citenamefont {Henriquez}, \citenamefont {Garc\'{\i}a},
  \citenamefont {Dumas}, \citenamefont {Bozzo},\ and\ \citenamefont
  {Pomar}}]{Gonzalez-Fuentes21}%
  \BibitemOpen
  \bibfield  {author} {\bibinfo {author} {\bibfnamefont {C.}~\bibnamefont
  {Gonzalez-Fuentes}}, \bibinfo {author} {\bibfnamefont {R.}~\bibnamefont
  {Henriquez}}, \bibinfo {author} {\bibfnamefont {C.}~\bibnamefont
  {Garc\'{\i}a}}, \bibinfo {author} {\bibfnamefont {R.~K.}\ \bibnamefont
  {Dumas}}, \bibinfo {author} {\bibfnamefont {B.}~\bibnamefont {Bozzo}},\ and\
  \bibinfo {author} {\bibfnamefont {A.}~\bibnamefont {Pomar}},\ }\href
  {https://doi.org/10.1103/PhysRevB.103.224403} {\bibfield  {journal} {\bibinfo
   {journal} {Phys. Rev. B}\ }\textbf {\bibinfo {volume} {103}},\ \bibinfo
  {pages} {224403} (\bibinfo {year} {2021})}\BibitemShut {NoStop}%
\bibitem [{\citenamefont {Foros}\ \emph {et~al.}(2005)\citenamefont {Foros},
  \citenamefont {Woltersdorf}, \citenamefont {Heinrich},\ and\ \citenamefont
  {Brataas}}]{Foros05}%
  \BibitemOpen
  \bibfield  {author} {\bibinfo {author} {\bibfnamefont {J.}~\bibnamefont
  {Foros}}, \bibinfo {author} {\bibfnamefont {G.}~\bibnamefont {Woltersdorf}},
  \bibinfo {author} {\bibfnamefont {B.}~\bibnamefont {Heinrich}},\ and\
  \bibinfo {author} {\bibfnamefont {A.}~\bibnamefont {Brataas}},\ }\href
  {https://doi.org/10.1063/1.1853131} {\bibfield  {journal} {\bibinfo
  {journal} {Journal of Applied Physics}\ }\textbf {\bibinfo {volume} {97}},\
  \bibinfo {pages} {10A714} (\bibinfo {year} {2005})}\BibitemShut {NoStop}%
\bibitem [{\citenamefont {Allen}\ \emph {et~al.}(2015)\citenamefont {Allen},
  \citenamefont {Manipatruni}, \citenamefont {Nikonov}, \citenamefont {Doczy},\
  and\ \citenamefont {Young}}]{Allen15}%
  \BibitemOpen
  \bibfield  {author} {\bibinfo {author} {\bibfnamefont {G.}~\bibnamefont
  {Allen}}, \bibinfo {author} {\bibfnamefont {S.}~\bibnamefont {Manipatruni}},
  \bibinfo {author} {\bibfnamefont {D.~E.}\ \bibnamefont {Nikonov}}, \bibinfo
  {author} {\bibfnamefont {M.}~\bibnamefont {Doczy}},\ and\ \bibinfo {author}
  {\bibfnamefont {I.~A.}\ \bibnamefont {Young}},\ }\href
  {https://doi.org/10.1103/PhysRevB.91.144412} {\bibfield  {journal} {\bibinfo
  {journal} {Phys. Rev. B}\ }\textbf {\bibinfo {volume} {91}},\ \bibinfo
  {pages} {144412} (\bibinfo {year} {2015})}\BibitemShut {NoStop}%
\bibitem [{\citenamefont {Shashank}\ \emph {et~al.}(2021)\citenamefont
  {Shashank}, \citenamefont {Medwal}, \citenamefont {Shibata}, \citenamefont
  {Nongjai}, \citenamefont {Vas}, \citenamefont {Duchamp}, \citenamefont
  {Asokan}, \citenamefont {Rawat}, \citenamefont {Asada}, \citenamefont
  {Gupta},\ and\ \citenamefont {Fukuma}}]{Shashank21}%
  \BibitemOpen
  \bibfield  {author} {\bibinfo {author} {\bibfnamefont {U.}~\bibnamefont
  {Shashank}}, \bibinfo {author} {\bibfnamefont {R.}~\bibnamefont {Medwal}},
  \bibinfo {author} {\bibfnamefont {T.}~\bibnamefont {Shibata}}, \bibinfo
  {author} {\bibfnamefont {R.}~\bibnamefont {Nongjai}}, \bibinfo {author}
  {\bibfnamefont {J.~V.}\ \bibnamefont {Vas}}, \bibinfo {author} {\bibfnamefont
  {M.}~\bibnamefont {Duchamp}}, \bibinfo {author} {\bibfnamefont
  {K.}~\bibnamefont {Asokan}}, \bibinfo {author} {\bibfnamefont {R.~S.}\
  \bibnamefont {Rawat}}, \bibinfo {author} {\bibfnamefont {H.}~\bibnamefont
  {Asada}}, \bibinfo {author} {\bibfnamefont {S.}~\bibnamefont {Gupta}},\ and\
  \bibinfo {author} {\bibfnamefont {Y.}~\bibnamefont {Fukuma}},\ }\href
  {https://doi.org/10.1002/qute.202000112} {\bibfield  {journal} {\bibinfo
  {journal} {Advanced Quantum Technologies}\ }\textbf {\bibinfo {volume} {4}},\
  \bibinfo {pages} {2000112} (\bibinfo {year} {2021})}\BibitemShut {NoStop}%
\bibitem [{\citenamefont {Karimeddiny}\ and\ \citenamefont
  {Ralph}(2021)}]{Karimeddiny21}%
  \BibitemOpen
  \bibfield  {author} {\bibinfo {author} {\bibfnamefont {S.}~\bibnamefont
  {Karimeddiny}}\ and\ \bibinfo {author} {\bibfnamefont {D.~C.}\ \bibnamefont
  {Ralph}},\ }\href {https://doi.org/10.1103/PhysRevApplied.15.064017}
  {\bibfield  {journal} {\bibinfo  {journal} {Phys. Rev. Applied}\ }\textbf
  {\bibinfo {volume} {15}},\ \bibinfo {pages} {064017} (\bibinfo {year}
  {2021})}\BibitemShut {NoStop}%
\bibitem [{\citenamefont {Liu}\ \emph {et~al.}(2012)\citenamefont {Liu},
  \citenamefont {Pai}, \citenamefont {Li}, \citenamefont {Tseng}, \citenamefont
  {Ralph},\ and\ \citenamefont {Buhrman}}]{Liu12}%
  \BibitemOpen
  \bibfield  {author} {\bibinfo {author} {\bibfnamefont {L.}~\bibnamefont
  {Liu}}, \bibinfo {author} {\bibfnamefont {C.-F.}\ \bibnamefont {Pai}},
  \bibinfo {author} {\bibfnamefont {Y.}~\bibnamefont {Li}}, \bibinfo {author}
  {\bibfnamefont {H.~W.}\ \bibnamefont {Tseng}}, \bibinfo {author}
  {\bibfnamefont {D.~C.}\ \bibnamefont {Ralph}},\ and\ \bibinfo {author}
  {\bibfnamefont {R.~A.}\ \bibnamefont {Buhrman}},\ }\href
  {https://doi.org/10.1126/science.1218197} {\bibfield  {journal} {\bibinfo
  {journal} {Science}\ }\textbf {\bibinfo {volume} {336}},\ \bibinfo {pages}
  {555} (\bibinfo {year} {2012})}\BibitemShut {NoStop}%
\bibitem [{\citenamefont {Huang}\ \emph {et~al.}(2018)\citenamefont {Huang},
  \citenamefont {He}, \citenamefont {Yap},\ and\ \citenamefont
  {Lim}}]{Huang18}%
  \BibitemOpen
  \bibfield  {author} {\bibinfo {author} {\bibfnamefont {L.}~\bibnamefont
  {Huang}}, \bibinfo {author} {\bibfnamefont {S.}~\bibnamefont {He}}, \bibinfo
  {author} {\bibfnamefont {Q.~J.}\ \bibnamefont {Yap}},\ and\ \bibinfo {author}
  {\bibfnamefont {S.~T.}\ \bibnamefont {Lim}},\ }\href
  {https://doi.org/10.1063/1.5036836} {\bibfield  {journal} {\bibinfo
  {journal} {Applied Physics Letters}\ }\textbf {\bibinfo {volume} {113}},\
  \bibinfo {pages} {022402} (\bibinfo {year} {2018})},\ \Eprint
  {https://arxiv.org/abs/https://doi.org/10.1063/1.5036836}
  {https://doi.org/10.1063/1.5036836} \BibitemShut {NoStop}%
\bibitem [{\citenamefont {Kondou}\ \emph {et~al.}(2012)\citenamefont {Kondou},
  \citenamefont {Sukegawa}, \citenamefont {Mitani}, \citenamefont
  {Tsukagoshi},\ and\ \citenamefont {Kasai}}]{Kondou12}%
  \BibitemOpen
  \bibfield  {author} {\bibinfo {author} {\bibfnamefont {K.}~\bibnamefont
  {Kondou}}, \bibinfo {author} {\bibfnamefont {H.}~\bibnamefont {Sukegawa}},
  \bibinfo {author} {\bibfnamefont {S.}~\bibnamefont {Mitani}}, \bibinfo
  {author} {\bibfnamefont {K.}~\bibnamefont {Tsukagoshi}},\ and\ \bibinfo
  {author} {\bibfnamefont {S.}~\bibnamefont {Kasai}},\ }\href
  {https://doi.org/10.1143/apex.5.073002} {\bibfield  {journal} {\bibinfo
  {journal} {Applied Physics Express}\ }\textbf {\bibinfo {volume} {5}},\
  \bibinfo {pages} {073002} (\bibinfo {year} {2012})}\BibitemShut {NoStop}%
\bibitem [{\citenamefont {Tao}\ \emph {et~al.}(2018)\citenamefont {Tao},
  \citenamefont {Liu}, \citenamefont {Miao}, \citenamefont {Yu}, \citenamefont
  {Feng}, \citenamefont {Sun}, \citenamefont {You}, \citenamefont {Du},
  \citenamefont {Chen}, \citenamefont {Zhang}, \citenamefont {Zhang},
  \citenamefont {Yuan}, \citenamefont {Wu},\ and\ \citenamefont
  {Ding}}]{Tao18}%
  \BibitemOpen
  \bibfield  {author} {\bibinfo {author} {\bibfnamefont {X.}~\bibnamefont
  {Tao}}, \bibinfo {author} {\bibfnamefont {Q.}~\bibnamefont {Liu}}, \bibinfo
  {author} {\bibfnamefont {B.}~\bibnamefont {Miao}}, \bibinfo {author}
  {\bibfnamefont {R.}~\bibnamefont {Yu}}, \bibinfo {author} {\bibfnamefont
  {Z.}~\bibnamefont {Feng}}, \bibinfo {author} {\bibfnamefont {L.}~\bibnamefont
  {Sun}}, \bibinfo {author} {\bibfnamefont {B.}~\bibnamefont {You}}, \bibinfo
  {author} {\bibfnamefont {J.}~\bibnamefont {Du}}, \bibinfo {author}
  {\bibfnamefont {K.}~\bibnamefont {Chen}}, \bibinfo {author} {\bibfnamefont
  {S.}~\bibnamefont {Zhang}}, \bibinfo {author} {\bibfnamefont
  {L.}~\bibnamefont {Zhang}}, \bibinfo {author} {\bibfnamefont
  {Z.}~\bibnamefont {Yuan}}, \bibinfo {author} {\bibfnamefont {D.}~\bibnamefont
  {Wu}},\ and\ \bibinfo {author} {\bibfnamefont {H.}~\bibnamefont {Ding}},\
  }\bibfield  {journal} {\bibinfo  {journal} {Science Advances}\ }\textbf
  {\bibinfo {volume} {4}},\ \href {https://doi.org/10.1126/sciadv.aat1670}
  {10.1126/sciadv.aat1670} (\bibinfo {year} {2018})\BibitemShut {NoStop}%
\bibitem [{\citenamefont {Fan}\ \emph {et~al.}(2014)\citenamefont {Fan},
  \citenamefont {Celik}, \citenamefont {Wu}, \citenamefont {Ni}, \citenamefont
  {Lee}, \citenamefont {Lorenz},\ and\ \citenamefont {Xiao}}]{Fan14}%
  \BibitemOpen
  \bibfield  {author} {\bibinfo {author} {\bibfnamefont {X.}~\bibnamefont
  {Fan}}, \bibinfo {author} {\bibfnamefont {H.}~\bibnamefont {Celik}}, \bibinfo
  {author} {\bibfnamefont {J.}~\bibnamefont {Wu}}, \bibinfo {author}
  {\bibfnamefont {C.}~\bibnamefont {Ni}}, \bibinfo {author} {\bibfnamefont
  {K.-J.}\ \bibnamefont {Lee}}, \bibinfo {author} {\bibfnamefont {V.~O.}\
  \bibnamefont {Lorenz}},\ and\ \bibinfo {author} {\bibfnamefont {J.~Q.}\
  \bibnamefont {Xiao}},\ }\href {https://doi.org/10.1038/ncomms4042} {\bibfield
   {journal} {\bibinfo  {journal} {Nature Communications}\ }\textbf {\bibinfo
  {volume} {5}},\ \bibinfo {pages} {3042} (\bibinfo {year} {2014})}\BibitemShut
  {NoStop}%
\bibitem [{\citenamefont {Rojas-S\'anchez}\ \emph {et~al.}(2014)\citenamefont
  {Rojas-S\'anchez}, \citenamefont {Reyren}, \citenamefont {Laczkowski},
  \citenamefont {Savero}, \citenamefont {Attan\'e}, \citenamefont {Deranlot},
  \citenamefont {Jamet}, \citenamefont {George}, \citenamefont {Vila},\ and\
  \citenamefont {Jaffr\`es}}]{Rojas-Sanchez14}%
  \BibitemOpen
  \bibfield  {author} {\bibinfo {author} {\bibfnamefont {J.-C.}\ \bibnamefont
  {Rojas-S\'anchez}}, \bibinfo {author} {\bibfnamefont {N.}~\bibnamefont
  {Reyren}}, \bibinfo {author} {\bibfnamefont {P.}~\bibnamefont {Laczkowski}},
  \bibinfo {author} {\bibfnamefont {W.}~\bibnamefont {Savero}}, \bibinfo
  {author} {\bibfnamefont {J.-P.}\ \bibnamefont {Attan\'e}}, \bibinfo {author}
  {\bibfnamefont {C.}~\bibnamefont {Deranlot}}, \bibinfo {author}
  {\bibfnamefont {M.}~\bibnamefont {Jamet}}, \bibinfo {author} {\bibfnamefont
  {J.-M.}\ \bibnamefont {George}}, \bibinfo {author} {\bibfnamefont
  {L.}~\bibnamefont {Vila}},\ and\ \bibinfo {author} {\bibfnamefont
  {H.}~\bibnamefont {Jaffr\`es}},\ }\href
  {https://doi.org/10.1103/PhysRevLett.112.106602} {\bibfield  {journal}
  {\bibinfo  {journal} {Phys. Rev. Lett.}\ }\textbf {\bibinfo {volume} {112}},\
  \bibinfo {pages} {106602} (\bibinfo {year} {2014})}\BibitemShut {NoStop}%
\bibitem [{\citenamefont {Sagasta}\ \emph {et~al.}(2016)\citenamefont
  {Sagasta}, \citenamefont {Omori}, \citenamefont {Isasa}, \citenamefont
  {Gradhand}, \citenamefont {Hueso}, \citenamefont {Niimi}, \citenamefont
  {Otani},\ and\ \citenamefont {Casanova}}]{Sagasta16}%
  \BibitemOpen
  \bibfield  {author} {\bibinfo {author} {\bibfnamefont {E.}~\bibnamefont
  {Sagasta}}, \bibinfo {author} {\bibfnamefont {Y.}~\bibnamefont {Omori}},
  \bibinfo {author} {\bibfnamefont {M.}~\bibnamefont {Isasa}}, \bibinfo
  {author} {\bibfnamefont {M.}~\bibnamefont {Gradhand}}, \bibinfo {author}
  {\bibfnamefont {L.~E.}\ \bibnamefont {Hueso}}, \bibinfo {author}
  {\bibfnamefont {Y.}~\bibnamefont {Niimi}}, \bibinfo {author} {\bibfnamefont
  {Y.}~\bibnamefont {Otani}},\ and\ \bibinfo {author} {\bibfnamefont
  {F.}~\bibnamefont {Casanova}},\ }\href
  {https://doi.org/10.1103/PhysRevB.94.060412} {\bibfield  {journal} {\bibinfo
  {journal} {Phys. Rev. B}\ }\textbf {\bibinfo {volume} {94}},\ \bibinfo
  {pages} {060412} (\bibinfo {year} {2016})}\BibitemShut {NoStop}%
\bibitem [{\citenamefont {Zhang}\ \emph {et~al.}(2002)\citenamefont {Zhang},
  \citenamefont {Levy},\ and\ \citenamefont {Fert}}]{Zhang02}%
  \BibitemOpen
  \bibfield  {author} {\bibinfo {author} {\bibfnamefont {S.}~\bibnamefont
  {Zhang}}, \bibinfo {author} {\bibfnamefont {P.~M.}\ \bibnamefont {Levy}},\
  and\ \bibinfo {author} {\bibfnamefont {A.}~\bibnamefont {Fert}},\ }\href
  {https://doi.org/10.1103/PhysRevLett.88.236601} {\bibfield  {journal}
  {\bibinfo  {journal} {Phys. Rev. Lett.}\ }\textbf {\bibinfo {volume} {88}},\
  \bibinfo {pages} {236601} (\bibinfo {year} {2002})}\BibitemShut {NoStop}%
\bibitem [{\citenamefont {Manchon}(2012)}]{Manchon12}%
  \BibitemOpen
  \bibfield  {author} {\bibinfo {author} {\bibfnamefont {A.}~\bibnamefont
  {Manchon}},\ }\href@noop {} {\bibinfo {title} {Spin hall effect versus rashba
  torque: a diffusive approach}} (\bibinfo {year} {2012}),\ \Eprint
  {https://arxiv.org/abs/1204.4869} {arXiv:1204.4869 [cond-mat.mes-hall]}
  \BibitemShut {NoStop}%
\bibitem [{\citenamefont {Nan}\ \emph {et~al.}(2015)\citenamefont {Nan},
  \citenamefont {Emori}, \citenamefont {Boone}, \citenamefont {Wang},
  \citenamefont {Oxholm}, \citenamefont {Jones}, \citenamefont {Howe},
  \citenamefont {Brown},\ and\ \citenamefont {Sun}}]{Nan15}%
  \BibitemOpen
  \bibfield  {author} {\bibinfo {author} {\bibfnamefont {T.}~\bibnamefont
  {Nan}}, \bibinfo {author} {\bibfnamefont {S.}~\bibnamefont {Emori}}, \bibinfo
  {author} {\bibfnamefont {C.~T.}\ \bibnamefont {Boone}}, \bibinfo {author}
  {\bibfnamefont {X.}~\bibnamefont {Wang}}, \bibinfo {author} {\bibfnamefont
  {T.~M.}\ \bibnamefont {Oxholm}}, \bibinfo {author} {\bibfnamefont {J.~G.}\
  \bibnamefont {Jones}}, \bibinfo {author} {\bibfnamefont {B.~M.}\ \bibnamefont
  {Howe}}, \bibinfo {author} {\bibfnamefont {G.~J.}\ \bibnamefont {Brown}},\
  and\ \bibinfo {author} {\bibfnamefont {N.~X.}\ \bibnamefont {Sun}},\ }\href
  {https://doi.org/10.1103/PhysRevB.91.214416} {\bibfield  {journal} {\bibinfo
  {journal} {Phys. Rev. B}\ }\textbf {\bibinfo {volume} {91}},\ \bibinfo
  {pages} {214416} (\bibinfo {year} {2015})}\BibitemShut {NoStop}%
\bibitem [{\citenamefont {Haku}\ \emph {et~al.}(2020)\citenamefont {Haku},
  \citenamefont {Musha}, \citenamefont {Gao},\ and\ \citenamefont
  {Ando}}]{Haku20}%
  \BibitemOpen
  \bibfield  {author} {\bibinfo {author} {\bibfnamefont {S.}~\bibnamefont
  {Haku}}, \bibinfo {author} {\bibfnamefont {A.}~\bibnamefont {Musha}},
  \bibinfo {author} {\bibfnamefont {T.}~\bibnamefont {Gao}},\ and\ \bibinfo
  {author} {\bibfnamefont {K.}~\bibnamefont {Ando}},\ }\href
  {https://doi.org/10.1103/PhysRevB.102.024405} {\bibfield  {journal} {\bibinfo
   {journal} {Phys. Rev. B}\ }\textbf {\bibinfo {volume} {102}},\ \bibinfo
  {pages} {024405} (\bibinfo {year} {2020})}\BibitemShut {NoStop}%
\bibitem [{\citenamefont {Ando}\ and\ \citenamefont {Saitoh}(2010)}]{Ando10}%
  \BibitemOpen
  \bibfield  {author} {\bibinfo {author} {\bibfnamefont {K.}~\bibnamefont
  {Ando}}\ and\ \bibinfo {author} {\bibfnamefont {E.}~\bibnamefont {Saitoh}},\
  }\href {https://doi.org/10.1063/1.3517131} {\bibfield  {journal} {\bibinfo
  {journal} {Journal of Applied Physics}\ }\textbf {\bibinfo {volume} {108}},\
  \bibinfo {pages} {113925} (\bibinfo {year} {2010})},\ \Eprint
  {https://arxiv.org/abs/https://pubs.aip.org/aip/jap/article-pdf/doi/10.1063/1.3517131/15064900/113925\_1\_online.pdf}
  {https://pubs.aip.org/aip/jap/article-pdf/doi/10.1063/1.3517131/15064900/113925\_1\_online.pdf}
  \BibitemShut {NoStop}%
\end{thebibliography}%

\end{document}